%% file: main.tex
\documentclass[12pt]{article}

\usepackage{physics}
\usepackage[printonlyused, nolist]{acronym}
\usepackage{svg}
\usepackage{subcaption}
\usepackage[colorlinks,citecolor=blue,linkcolor=blue,urlcolor=blue,bookmarks=false,hypertexnames=true]{hyperref}
\usepackage{scalerel}
\usepackage{tikz}
\usetikzlibrary{svg.path}
\usepackage[style=numeric, sorting=none, sortcites=true]{biblatex}
\begin{document}

\include{command_definitions}
    
\title{\vspace{-3.0cm}\textbf{\Large Truncated Variational Hamiltonian Ansatz: efficient quantum circuit design for quantum chemistry and material science}}

\author{\normalsize
Clemens Possel\orcidicon{0000-0002-1859-7533}$^{1,2}$, 
Walter Hahn\orcidicon{0000-0003-4355-2019}$^3$, 
Reza Shirazi\orcidicon{0009-0000-7622-3161}$^4$,
Marina Walt\orcidicon{0000-0002-4147-9420}$^4$, \\ \normalsize
Peter Pinski\orcidicon{0000-0003-2874-6843}$^4$,
Frank K. Wilhelm\orcidicon{0000-0003-1034-8476}$^{5,2}$
and Dmitry Bagrets\orcidicon{0000-0002-3985-4834}$^{5}$
}

\date{\footnotesize
    $^1$ Fraunhofer Institute for Chemical Technology ICT, Joseph-von-Fraunhofer-Str. 7, 76327 Pfinztal, Germany \\
    $^2$ Department of Physics, Saarland University, 66123 Saarbr\"ucken, Germany \\
    $^3$ Fraunhofer Institute for Applied Solid State Physics IAF, Tullastr. 72, 79108 Freiburg, Germany \\
    $^4$ HQS Quantum Simulations GmbH, Rintheimer Straße 23, 76131 Karlsruhe, Germany \\
    $^5$ Peter Gr\"unberg Institute, Quantum Computing Analytics (PGI-12), Forschungszentrum J\"ulich, 52425 J\"ulich, Germany \\ [2ex]
    E-mail: clemens.possel@ict.fraunhofer.de
}

\maketitle

\section*{{\small Abstract}}
{\small
Quantum computing has the potential to revolutionize quantum chemistry and material science by offering solutions to complex problems unattainable with classical computers.
However, the development of efficient quantum algorithms that are efficient under noisy conditions remains a major challenge.
This paper introduces the \acf{tvha}, a novel circuit design for conducting quantum calculations on \acf{nisq} devices.
\ac{tvha} provides a promising approach for a broad range of applications by utilizing principles from the adiabatic theorem in solid state physics.
Our proposed ansatz significantly reduces the parameter count and can decrease circuit size substantially, with a trade-off in accuracy. Thus, \ac{tvha} facilitates easier convergence within the variational quantum eigensolver framework compared to state-of-the-art ansätze such as \ac{ucc} and \ac{hea}.
While this paper concentrates on the practical applications of \ac{tvha} in quantum chemistry, demonstrating its suitability for both weakly and strongly correlated systems and its compatibility with active space calculations, its underlying principles suggest a wider applicability extending to the broader field of material science computations on quantum computing platforms.}
\acresetall

\newpage

\input{1_section_introduction.tex}

\input{2_section_theory.tex}

\input{3_section_truncation_scheme_for_variational_hamiltonian_ansatz.tex}

\input{4_section_results_and_discussion.tex}

\input{5_section_summary_and_outlook}

\input{acronyms}

\input{acknowledgements}

\input{conflict_of_interests}

\printbibliography

\input{supplementary_information}

\end{document}

%% file: command_definitions.tex
\definecolor{orcidlogocol}{HTML}{A6CE39}
\tikzset{
  orcidlogo/.pic={
    \fill[orcidlogocol] svg{M256,128c0,70.7-57.3,128-128,128C57.3,256,0,198.7,0,128C0,57.3,57.3,0,128,0C198.7,0,256,57.3,256,128z};
    \fill[white] svg{M86.3,186.2H70.9V79.1h15.4v48.4V186.2z}
                 svg{M108.9,79.1h41.6c39.6,0,57,28.3,57,53.6c0,27.5-21.5,53.6-56.8,53.6h-41.8V79.1z M124.3,172.4h24.5c34.9,0,42.9-26.5,42.9-39.7c0-21.5-13.7-39.7-43.7-39.7h-23.7V172.4z}
                 svg{M88.7,56.8c0,5.5-4.5,10.1-10.1,10.1c-5.6,0-10.1-4.6-10.1-10.1c0-5.6,4.5-10.1,10.1-10.1C84.2,46.7,88.7,51.3,88.7,56.8z};
  }
}
\newcommand\orcidicon[1]{\href{https://orcid.org/#1}{\mbox{\scalerel*{
\begin{tikzpicture}[yscale=-1,transform shape]
\pic{orcidlogo};
\end{tikzpicture}
}{|}}}}

\newcommand{\eref}[1]{(\ref{#1})}
\newcommand{\sref}[1]{section~\ref{#1}}
\newcommand{\fref}[1]{figure~\ref{#1}}
\newcommand{\tref}[1]{table~\ref{#1}}
\newcommand{\Eref}[1]{Equation (\ref{#1})}
\newcommand{\Sref}[1]{Section~\ref{#1}}
\newcommand{\Fref}[1]{Figure~\ref{#1}}
\newcommand{\Tref}[1]{Table~\ref{#1}}

\setcounter{topnumber}{2}
\setcounter{bottomnumber}{2}
\setcounter{totalnumber}{4}
\renewcommand{\topfraction}{0.85}
\renewcommand{\bottomfraction}{0.85}
\renewcommand{\textfraction}{0.15}
\renewcommand{\floatpagefraction}{0.7}

%% file: 1_section_introduction.tex
\section{Introduction}

Quantum computing has the potential to solve complex problems beyond the reach of classical computers.
A prime candidate for this transformation is quantum chemistry, particularly due to the unfavorable scaling of computational methods that solve the Schrödinger equation exactly on classical computing hardware. Numerous efforts have been made to develop approaches for applying quantum computers to quantum chemistry, primarily through algorithms based on the variational principle, to leverage \ac{nisq} devices in this domain\cite{ALEXEEV2024666,bharti_noisy_2022,tilly_variational_2022,cerezo_variational_2021}.

An exact solution to the Schrödinger equation---within the limitations of the atomic basis set---is provided by the \ac{fci} method\cite{SzaboOstlund,Helgaker_Book_2000}.
Its computational cost scales exponentially with the system size, thus rendering it unfeasible for any but the smallest systems.
Approximations to the \ac{fci} method have been developed through approaches such as \ac{dmrg} or various stochastic and deterministic selected configuration interaction schemes.
While these can push the boundaries of such calculations on classical computers, the scaling wall still sets a stiff limit to their general applicability.
We refer the interested reader to review articles for further information on that topic\cite{Eriksen_2020,Baiardi_2020}.
Thus, computational chemistry is approached on classical computers with approximations that lead to polynomial, rather than exponential scaling of the computational cost with system size.
A conceptual starting point for many approaches is \ac{hf} theory: the wave function is approximated through a Slater determinant, and the orbitals are adapted to minimize the energy expection value\cite{SzaboOstlund,Helgaker_Book_2000}.
It is a mean-field theory, and the difference to the exact (\ac{fci}) solution is referred to as the correlation energy.
In practice, computational chemistry often resorts to \ac{dft} (we suggest~\cite{Bursch_2022} for practical recommendations).
A more systematic approach to recover correlation energy is to employ dynamic correlation methods, such as truncated \ac{ci}, \ac{mp} Perturbation Theory or \ac{cc}\cite{SzaboOstlund,Helgaker_Book_2000}.
Local correlation techniques have reached a level of maturity that permits approximate \ac{cc} calculations, particularly for single-point energies, to be performed routinely for systems with tens or even hundreds atoms, though derivatives and properties remain bigger challenges\cite{Ma_2018,Riplinger_CCSD_2013,Riplinger_CCSDT_2013}.

In quantum chemistry, dynamic correlation is distinguished conceptually from static correlation.
While dynamic correlation encompasses relatively small contributions from many electrons, static correlation refers specifically to strongly correlated electrons.
In practice, the transition between these two types is rather smooth, but the methods employed to treat both types of correlation are very different.
For static correlation, the dominant approach are complete active space methods: a set of orbitals, the active space, is selected for a \ac{fci} treatment only within that subspace.
The remaining orbitals, and their interaction with the active space, are left at a mean-field level.
Popular examples are the \ac{casscf} and \ac{casci} methods\cite{Roos_Book_2016}.
Active space methods have received interest as a potential application domain of quantum computing, as they provide a partitioning of the system where the quantum computer would substitute for the \ac{ci} calculation on a classical computer---a more accessible goal than attempting to treat all orbitals and electrons in a molecule at once\cite{Bauer_2020,Takeshita_2020}.

Given the limitations of classical approaches in addressing dynamic and static correlation effectively, there is a growing interest in exploring quantum algorithms that can offer significant advantages in computational efficiency and accuracy.
In this context, fault-tolerant quantum algorithms such as the \ac{qpe} have emerged as promising candidates, leveraging the unique capabilities of quantum computers to manage complex correlations in quantum systems more effectively than their classical counterparts.
The \ac{qpe} algorithm~\cite{nielsen2010quantum}, while providing guaranteed algorithmic speed-up, generates large circuits with hundreds of thousands of gates that far exceed the capacity of current \ac{nisq} devices, which can only manage circuits with at most few hundred gates.
In response to these limitations, \acp{vqa}\cite{Peruzzo.2014,tilly_variational_2022,ALEXEEV2024666} have been developed to provide more compact circuits, though they require classical optimization of variational parameters.
Until recently, the primary methods employed in quantum chemistry calculations on quantum computers included \ac{uccsd}\cite{barkoutsos2018ucc} and \ac{hea}\cite{kandala2017hea,leone2022hea}, both of which present distinct drawbacks.
\ac{uccsd}, while rooted in quantum chemistry, frequently results in large circuits with numerous parameters, making it challenging to implement on \ac{nisq} devices due to their limited coherence times and gate fidelity.
Conversely, \ac{hea}, designed to minimize circuit depth, may lack the necessary precision and systematic improvement path for more accurate quantum simulations, as it does not inherently incorporate the physical properties of the molecular systems.
Derived from solid state physics, \ac{tvha} proposed in this paper addresses some of these issues by reducing the parameter count compared to \ac{uccsd} and hardware-efficient approaches like \ac{hea}, while still encountering circuit sizes comparable to those of \ac{uccsd}.
\ac{tvha} introduces a novel truncation scheme that optimizes the operators used in circuit construction to further reduce circuit depth and outperform existing methods for the instances tested.
This approach strives to balance accuracy and efficiency, thus making quantum calculations more feasible on \ac{nisq} devices.

This paper is structured as follows: \Sref{sec:preliminary_considerations} outlines the theoretical foundations from quantum chemistry and solid state physics, followed by the development of the unitary operator for the \ac{tvha} ansatz and the construction of the quantum circuit in \sref{sec:truncation_scheme_for_vha}.
The subsequent \sref{sec:results_of_numerical_calculations} presents numerical simulations of the test systems \ac{lih}, \ac{h2}, \ac{h4}, and \ac{ch2}, assessing the ansatz's applicability for quantum chemistry systems.
\Sref{sec:summary_and_outlook} summarizes our findings and gives an outlook to possible future research.

%% file: 2_section_theory.tex
\section{Preliminary Considerations} \label{sec:preliminary_considerations}

This section forms the theoretical basis for the paper by exploring key concepts fundamental to the development of \ac{tvha}.
First, we present the adiabatic theorem for universal quantum computing, which forms the basis for the \ac{vha} algorithm.
This is followed by a detailed look at the quantum chemistry Hamiltonian with specific emphasis on its components as well as the \ac{hf} approximation--a classically tractable starting point for evolving the quantum system.
Lastly, the section examines the concept and methods of active space calculations in quantum chemistry, which have been utilized for our \ac{ch2} test system and are crucial to reach larger system sizes.
These fundamental concepts establish the basis for the development of \ac{tvha}.

\subsection{Adiabatic Theorem for Universal Quantum Computing}

The adiabatic theorem states that a system stays in its instantaneous eigenstate under transformation of an initial Hamiltonian $H_0$ to the final Hamiltonian $H_f$ in the limit of sufficiently slow transformation (see~\cite{albash2018adiabatic} for a comprehensive overview).
If the system is initialized in the ground state, this procedure can be used to determine the ground state of the final system.

Assuming a linear ramping function, that does not affect the condition for adiabaticity and provides a sufficient framework to interpolate between the Hamiltonians over time without favoring any particular time step, the Hamiltonian for the adiabatic time evolution can be written as
\begin{equation}
    H(\tau) = (1-\tau/T) H_0 + (\tau/T) H_f,\label{eq:adiabatic_hamiltonian}
\end{equation}
where $T$ is the total time of the adiabatic evolution which needs to be set sufficiently large to avoid level crossings~\cite{Farhi.2000}.
For non-vanishing energy gaps, the required time $T$ for adiabatic evolution depends on the energy gap such that the condition
\begin{equation}
    \left | \frac{\bra{\psi_1(\tau)} d_\tau {H(\tau)} \ket{\psi_0(\tau)}}{{\Delta E(\tau)}^2} \right | \ll 1 \qquad \forall \tau \in [0,T], \label{eq:adiabatic_energy_gap}
\end{equation}
must be satisfied with the energy gap $\Delta E(\tau)$ being defined as the energy difference between the ground state $\ket{\psi_0(\tau)}$ and the first excited state $\ket{\psi_1(\tau)}$ of the Hamiltonian $H(\tau)$.

In principle, adiabatic parameterizations other than the linear ramping used in \eref{eq:adiabatic_hamiltonian} can be used to turn on the interaction Hamiltonian, provided that $H(0)=H_0$ and $H(T)=H_f$ is fulfilled~\cite{Perdomo2011}.
Additionally, shortcuts to the adiabatic evolution, for example counterdiabatic driving, are discussed in the literature and could potentially be applied to the proposed \ac{tvha} method\cite{an2016counterdiabaticshortcuts}.

The wave function of the ground state of the final system is then formally given by
\begin{equation}
    \ket{\psi} = \mathcal{T} \exp[i \int_0^T H(\tau) d\tau] \ket{\psi_0},
    \label{eq:adiabatic_evolution_of_wavefunction}
\end{equation}
with $\ket{\psi_0}$ being the ground state of the initial system and $\mathcal{T}$ denoting the time ordering operator.

For gate-based quantum computers it is necessary to discretize the adiabatic evolution since their operations are executed through a sequence of discrete unitary gates rather than continuous time evolution governed by a smoothly changing Hamiltonian, which is the fundamental principle of quantum annealers.

\Eref{eq:adiabatic_evolution_of_wavefunction} can be approximated by
\begin{equation}
    \ket{\psi} = \prod_{n=1}^N \exp[i H(n \Delta \tau) \Delta \tau] \ket{\psi_0}, \qquad \Delta \tau = T/N
    \label{eq:discrete_adiabatic_evolution}
\end{equation}
where $N$ is the number of discretization steps with $\Delta \tau$ being the time step.
This equation of the discrete time evolution is exact in the limit of $\Delta \tau \rightarrow{0}$ (or equivalently $N \rightarrow \infty$), provided $T$ is sufficiently long so that the adiabaticity condition \eqref{eq:adiabatic_energy_gap}
is satisfied.

In general, the Hamiltonian $H(n \Delta\tau)$ consists of multiple non-commuting operators.
In order to evaluate these on gate-based quantum computers, Suzuki-Trotter expansion is usually applied which introduces an approximation error called Trotter error due to non-commuting terms in the Hamiltonian\cite{suzuki}.
In this paper, Suzuki-Trotter expansion is used in first order, i.e. $\exp(i (A  + B) ) \approx \exp(i A) \exp(i B)$, where $A$ and $B$ are non-commuting Pauli strings, $[A,B] \neq 0$, from the Hamiltonian $H(n \Delta\tau)$ of \eref{eq:discrete_adiabatic_evolution} after transformation to Pauli operators (in our case 
achieved via the Jordan-Wigner mapping of the fermionic many-body operators).
The unitaries $\exp(i A)$ and $\exp(i B)$ arising after Suzuki Trotter approximation can be expressed as quantum gates (see also \fref{fig:circuits_non_coulomb_two_body}).
Higher order expansions can further increase the accuracy of the approximation but result in highly increased circuit depths.
The error introduced by larger circuits of \ac{nisq} hardware is expected to be larger than the gain within the approximation error.
Also, the common expectation is that the Trotter error will be at least partially suppressed by the optimization procedure.

\subsection{Quantum Chemistry Hamiltonian} \label{subsec:quantum_chemistry_hamiltonian}

The quantum chemistry description of a molecule encompasses both the nuclei and their electrons. Utilizing the Born-Oppenheimer approximation allows for the separation of the degrees of freedom of the nuclei and electrons, enabling the electrons to be characterized within an external field generated by the nuclei. The Hamiltonian of this electronic system can be written in second quantization as
\begin{equation}
    H = \sum_{ij} h_{ij} a_i^\dagger a_j  + \frac{1}{2}  \sum_{ijk\ell} g_{ijk\ell} a_i^\dagger a_j^\dagger a_k a_\ell = H_{\text{one-body}} + H_{\text{two-body}}. \label{eq:quantum_chemistry_hamiltonian}
\end{equation}
The one-electron integrals $h_{ij}$ and two-electron integrals $g_{ijk\ell}$ can be pre-calculated on a classical computer using standard quantum chemistry calculation software such as \ac{pyscf}\cite{pyscf2018}.
By construction, the two-body matrix elements obey the symmetry relation $g_{ijk\ell} = g_{ji\ell k}$.

The first component of this Hamiltonian $H_{\text{one-body}}$ contains the one-body terms (commonly referred to as single excitations in chemical terminology), while the second part $H_{\text{two-body}}$ contains the less trivial two-body terms (double excitations) including Coulomb interaction terms, exchange terms and further terms.

The two-body terms $g_{ijk\ell} a_i^\dagger a_j^\dagger a_k a_\ell$ can be further decomposed into Coulomb terms and non-Coulomb terms.
The Coulomb terms are those, where either $i=\ell$ and $j=k$ or $i=k$ and $j=\ell$.
When, for instance, the first condition is satisfied, these terms can be written as 
\begin{equation}
    g_{ijji} a_i^\dagger a_j^\dagger a_j a_i = g_{ijji} a_i^\dagger a_i a_j^\dagger a_j = g_{ijji} \hat{n}_i \hat{n}_j,
    \label{eq:coulomb_terms}
\end{equation}
where $\hat{n}_i$ denotes the particle number operator of site $i$.
In this case, one can assume that $i\neq j$ because $(\hat{n}_i)^2 = \hat{n}_i$ and thus these terms are one-body terms.

All two-body terms that are not Coulomb terms (i.e. not of the form shown in \eref{eq:coulomb_terms}) are summarized as non-Coulomb two-body terms. Their magnitudes are the smallest for typical molecules, ranging from multiple orders smaller than the one-body and Coulomb two-body terms to sizes comparable to the Coulomb two-body terms  (compare e.g. with \fref{fig:hist_lih}).
In principle, one could further distinguish between terms where all indices $i$, $j$, $k$, $\ell$ are different and those where one pair of indices is the same, i.e. those with a single particle number operator $\hat{n}_i$.
Since their difference is negligible for our purposes, we disregard further distinction among these terms.

The number of one-body terms often scales linearly with the number of considered orbitals $m$; in general the number of one-body and Coulomb two-body terms can be estimated by $\mathcal{O}(m^2)$.
However, the number of non-Coulomb two-body terms is bounded by $\mathcal{O}(m^4)$, which usually far exceeds the number of all other terms.
Here, the distinction between molecular systems and those in solid-state physics becomes evident.
In solid-state systems, two-body terms are typically restricted to nearest neighbors or next-nearest neighbors and share the same prefactor.
This restriction allows for a further reduction in complexity, enabling the consideration of larger, often periodic, systems.

\subsubsection{Hartree-Fock Approximation}

Typically, \ac{hf} calculations serve as the foundation for more sophisticated methods, such as perturbation theory.
In this paper, the \ac{hf} solution is used as an initial guess for the system's ground state, which is subsequently evolved into the final fully interacting system.

Solving the eigenvalue problem is generally computationally demanding due to the presence of two-body terms, especially the non-Coulomb two-body terms.
The widely used \ac{hf} approximation addresses this by treating the two-body terms within a mean-field framework, thereby transforming the Hamiltonian into an effective one-body Hamiltonian and discarding electron correlation.
This non-correlated system is described by the Fock operator $\hat{F}$, which is defined by
\begin{equation}
   \hat{F} =  \sum_{ij} (h_{ij}  +  V_{ij}) a_i^\dagger a_j, \qquad V_{ij} = \sum_{k \in {\rm occ}} (g_{ikkj} - g_{ikjk}),
\end{equation}
where the sum in $V_{ij}$ only takes occupied orbitals $k$ into account.
Solutions of converged \ac{hf} \ac{scf} calculations are commonly represented via canonical molecular orbitals, which make the Fock operator diagonal:
\begin{equation}
   \hat{F} = \sum_i \varepsilon_i  a_i^\dagger a_i, 
\end{equation}
with $\{\varepsilon_i\}$ being a set of orbital energies.
The approximate eigenvalue problem is
\begin{equation}
    \hat{F} \ket{\psi_{HF}} = \sum_i \varepsilon_i \ket{\psi_{HF}}.
\end{equation}
The solution from this eigenvalue problem makes the Fock matrix block-diagonal, that is $F_{ov}=0$, where $o$ refers to occupied and $v$ to virtual orbital.
The resulting \ac{hf} wave function $\ket{\psi_{HF}}$ can be efficiently pre-calculated using classical computational resources and provides a good estimate for weakly correlated systems.
Thus, it serves as a good initial state for our subsequent quantum chemistry calculations.

\subsection{Active Space Calculations}

Static electron correlation is commonly treated with active space methods in quantum chemistry\cite{Roos_Book_2016}. It is usually possible to identify a subset of orbitals, the active space, that host particularly strongly correlated electrons.
The \ac{fci} problem is only solved inside the active space, while the rest of the system is subjected to a more feasible mean-field treatment.
The same reasoning can be applied to reduce the size and complexity of the Hilbert space for near-term quantum computers: only the active space is mapped onto the quantum computer, yielding smaller, more tractable circuits.
Note that it is usually still necessary to recover dynamic correlation on top of the complete active space calculations for quantitative accuracy.

A major disadvantage of active space methods is that the results depend on the choice of the active space, and results with incorrectly chosen orbitals can be meaningless.
Often this amounts to more of an art than a rigorous science, introducing both subjectivity and a hurdle towards the application of active space methods by non-specialists.
For example, it is necessary to include correlation partner orbitals. Likewise, including unimportant orbitals, leaving out important orbitals or breaking molecular symmetry can lead to unbalanced active spaces with undesired consequences\cite{Roos_Book_2016,Veryazov_2011}.
These aspects, discussed in the literature for classical active space methods, can be expected to remain if the classical \ac{ci} solver is replaced with a quantum computer.

Attempts have been made to make active space selection more systematic or even automatic\cite{Pulay_1988,reiher_stein,Sayfutyarova_2017,Khedkar_2019,activespaceselection2015keller}.
In this work, correct selection of the active space for methylene was supported by the Active Space Finder (version 1.0)~\cite{code_asf}.
The first step in the procedure of \ac{asf} is a \ac{mp2} calculation with an unrestricted \ac{hf} reference.
The natural orbitals of the unrelaxed \ac{mp2} density are used to select a subset of orbitals.
With this initial active space, an approximate but fast \ac{dmrg}-\ac{casci} calculation is performed.
Orbital entropies calculated from the \ac{dmrg} result quantify the overall extent of electron correlation for each orbital: with the help of an entropy threshold, the final set of active orbitals is chosen.
This orbital space is mapped on the quantum computer, while the remaining orbitals are subject to a classical treatment.

%% file: 3_section_truncation_scheme_for_variational_hamiltonian_ansatz.tex
\section{Truncation Scheme for \acl{vha}} \label{sec:truncation_scheme_for_vha}

This section is designed to present our truncation scheme for the \ac{vha}.
It begins with a recap of the \ac{vha}'s fundamental concept and its variational deviation from adiabatic evolution.
This is followed by a meticulous examination of the Hamiltonian's decomposition into its one-body and two-body components.
A brief note on parameter initialization precedes an in-depth discussion of the operators' contribution to the energy prediction and circuit size.
Building on these findings, the truncation scheme is then introduced, marking the central part of this paper.

\subsection{Variational Hamiltonian Ansatz}

The basic idea of \ac{vha} (first introduced by \citeauthor{Wecker.2015} as \ac{hva}~\cite{Wecker.2015}) is to compensate for the discretization error introduced from \eref{eq:adiabatic_evolution_of_wavefunction} to \eqref{eq:discrete_adiabatic_evolution} (as well as for the Trotterization error to some extend) by introducing a set of parameters $\{\theta_n\}$ such that
\begin{equation}
    \ket{\psi} = \mathcal{T} \prod_{n=1}^N \exp[i \theta_n H(n \Delta \tau) \Delta \tau] \ket{\psi_0}
    = \mathcal{T} \prod_{n=1}^N \exp[i \frac{\theta_n \tau}{N}  H(n \Delta \tau)] \ket{\psi_0}.
    \label{eq:vha}
\end{equation}
These parameters $\{\theta_n\}$ can be optimized with a \ac{vqe}\cite{Peruzzo.2014} routine in a hybrid way using a quantum computer to evaluate the average energy using the wave function from \eref{eq:vha} for a given set of parameters $\{\theta_n\}$ and a classical computer for the optimization routine, i.e. choosing the next set of parameters $\{\theta_n\}$.
For implementation it is handy to rescale the parameters $\{\theta_n\}$ via merging the factor $\tau/N$ into them.

\subsection{Hamiltonian Decomposition}

The number of free parameters $\{\theta_n\}$ in the variational time evolution described in \eref{eq:vha} equals the number of Trotter steps.
This choice of parameterization typically requires multiple Trotter steps for sufficient accuracy.
Thus, it is beneficial to introduce further parameters per step, allowing for fewer Trotter steps, i.e. shorter quantum circuits, although this is paid with a possibly increased number of parameters.
There are multiple choices to decompose the Hamiltonian and introduce additional parameters for each subset of the Hamiltonian.
The most obvious approach decomposes the Hamiltonian into one-body and two-body terms as in \eref{eq:quantum_chemistry_hamiltonian}, and assigns disjoint sets of parameters for them.
While this approach is straight forward, more sophisticated decomposition schemes allow to use more information from the underlying quantum system, for example a molecule, to gain better accuracy with smaller circuits.
In literature various decomposition schemes can be found; most are introduced in conjuction with the Hubbard model~\cite{Wecker.2015,reiner2019finding}, which describes interacting particles, typically electrons, on a lattice.
However, they are only partly extendable to molecules, because of the difference between the structure of their respective Hamiltonians.
For the decomposition scheme one can also get inspired by the conceptual somewhat similar \ac{qaoa}\cite{Wiersema.2020}, whose similarity becomes evident, when one interprets the cost Hamiltonian as one-body or non-interacting Hamiltonian and the mixer Hamiltonian as interacting Hamiltonian comprised of the two-body terms.
Within the decomposition scheme we propose in this paper, the idea is to take the underlying structure of molecular Hamiltonians into account and split the Hamiltonian into three parts.
The first part contains the one-body terms; this part is labeled $H_\alpha$ within the scope of the variational ansatz with the wave function from \eref{eq:vha}.
The two-body terms are decomposed into Coulomb and non-Coulomb two-body terms (as described in \sref{subsec:quantum_chemistry_hamiltonian}). The Hamiltonian containing the Coulomb two-body terms is labeled $H_\beta$, the Hamiltonian of the remaining non-Coulomb two-body terms $H_\gamma$.

With this decomposition scheme, the parameterized Hamiltonian from \eref{eq:vha} can now be adjusted, replacing the single set $\{\theta_n\}$ of variational parameters by the parameter sets $\{\alpha_n\}$, $\{\beta_n\}$, and $\{\gamma_n\}$ yielding the Hamiltonian for a given reduced time
\begin{equation}
    H(\tau) = H_\alpha(\tau) + H_\beta(\tau) + H_\gamma(\tau),
\end{equation}
where the time dependence is given by \eref{eq:adiabatic_hamiltonian}.
With the newly introduced variational parameters, \eref{eq:vha} becomes
\begin{equation}
    \ket{\psi} =  \mathcal{T} \prod_{n=1}^N \left\{
        \exp[i \alpha_n H_\alpha(n \Delta \tau)]
        \exp[i \beta_n H_\beta(n \Delta \tau)]
        \exp[i \gamma_n H_\gamma(n \Delta \tau)]
    \right\} \ket{\psi_0}.
    \label{eq:vha_decomposed}
\end{equation}
In principle, the order of the three exponentials is arbitrary within the scope of the applied first-order Trotterization; although it should be noted that the terms don't commute in general, so the ordering affects the Trotterization error that is introduced.
Since application of the Hamiltonian $H_\alpha$ onto the \ac{hf} state doesn't alter the state but at most adds a global phase $\phi$, i.e. $\exp[i \alpha_0 H_\alpha(0)] \ket{\psi_0} = \exp[i\phi] \ket{\psi_0}$, it is useful to apply first the two-body terms onto the \ac{hf} state and then the one-body terms on the now altered state.
In practical application this avoids the introduction of an unnecessary parameter as well as a possible noise source due to the useless quantum gates associated with the first layer of $H_\alpha$ (useless only if the term $H_\alpha$ were the first one to apply onto the \ac{hf} state).

\subsection{Parameter initialization}

For initialization of the variational parameters $\{\alpha_n\}$, $\{\beta_n\}$, and $\{\gamma_n\}$ it is natural to follow the adiabatic evolution as initial guess, as suggested in the work of \citeauthor{reiner2019finding}~\cite{reiner2019finding}.
The parameters for the one-body terms are set to $\alpha_n = 1$ for all $n$ (i.e. all times $\tau$).
$\{\beta_n\}$, and $\{\gamma_n\}$ follow a linear ramp from 0 to 1, i.e. $\beta_n = \frac{n}{N}$ and $\gamma_n = \frac{n}{N}$ with $n$ and $N$ defined in \eref{eq:discrete_adiabatic_evolution}.
"Randomized initialization of ansatz parameters is an alternative choice, allowing to find local minima far away from the adiabatic path.
However, in this paper we expect the adiabatic evolution as initial guess to be sufficiently good to find local minima within a given number of optimization iterations.
Also, it is reproducible which is considered being advantageous for scientific studies.

\subsection{Contribution of Operators to Quantum Circuit Size}

\begin{figure}[!htb]
    \centering
    \begin{subfigure}{0.48\textwidth}
        \centering
        \includegraphics[width=0.3\textwidth]{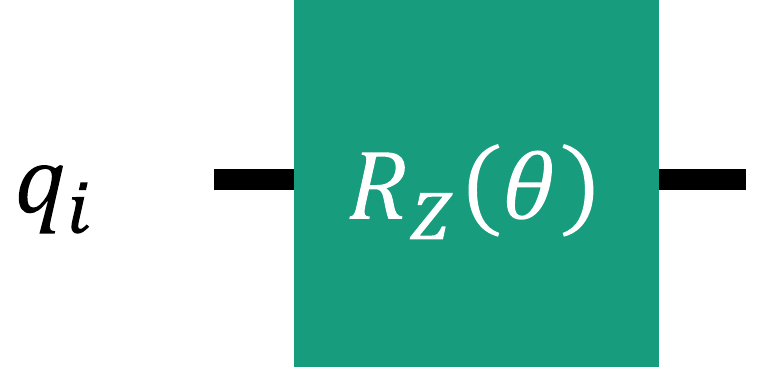}
        \caption{One-body term $h_{ii} a^\dagger_i a_i$ with the depicted parameter $\theta$ given by $\theta=-h_{ij}$}
        \label{fig:circuits_one_body}
    \end{subfigure}
    \begin{subfigure}{0.48\textwidth}
        \centering
        \includegraphics[width=0.9\textwidth]{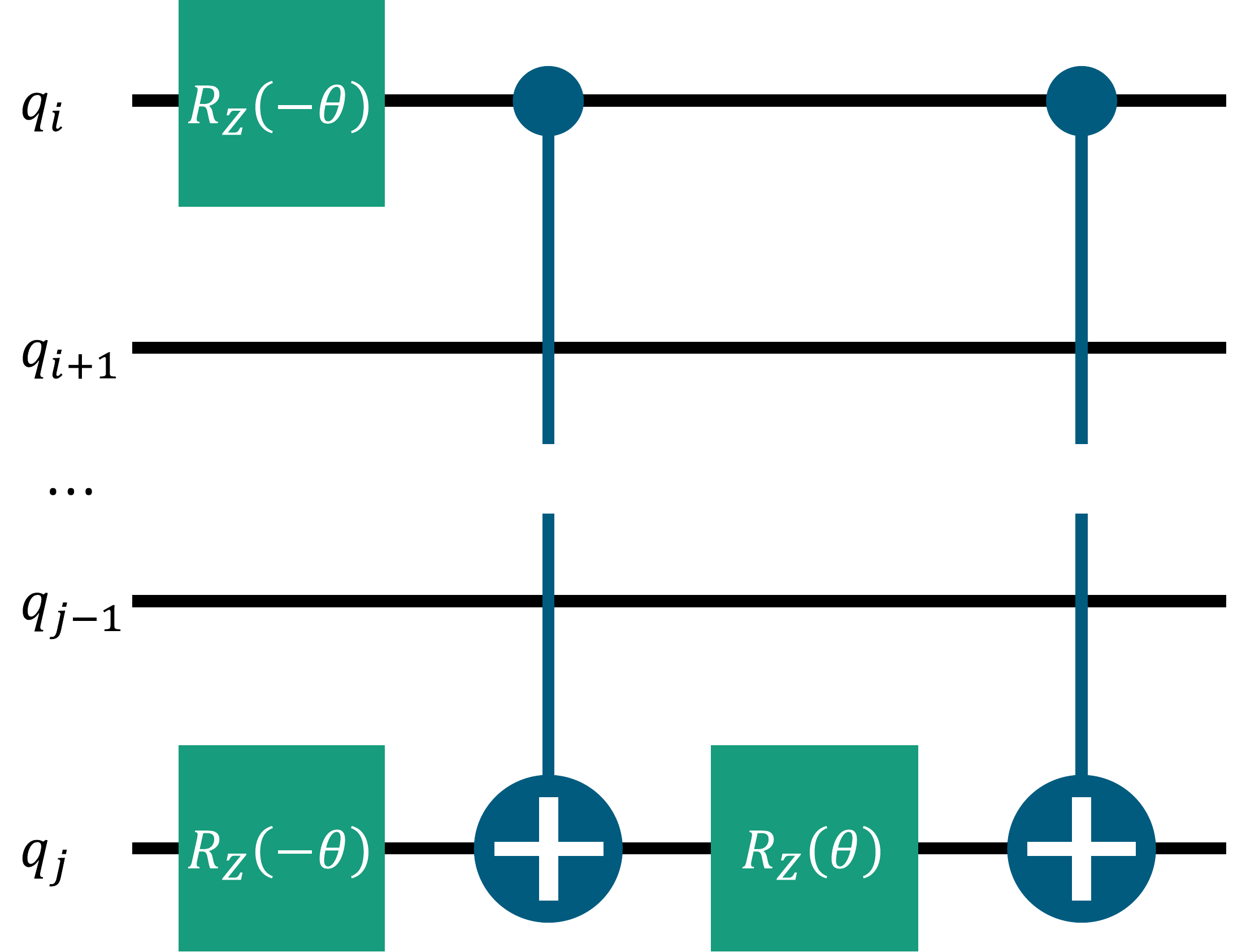}
        \caption{Coulomb two-body term $g_{ijji} a^\dagger_i a^\dagger_j a_j a_i$ with the parameter $\theta$ given by $\theta=g_{ijji}/2$}
        \label{fig:circuits_couloumb_two_body}
    \end{subfigure}
    \begin{subfigure}{\textwidth}
        \centering
        \includegraphics[width=\textwidth]{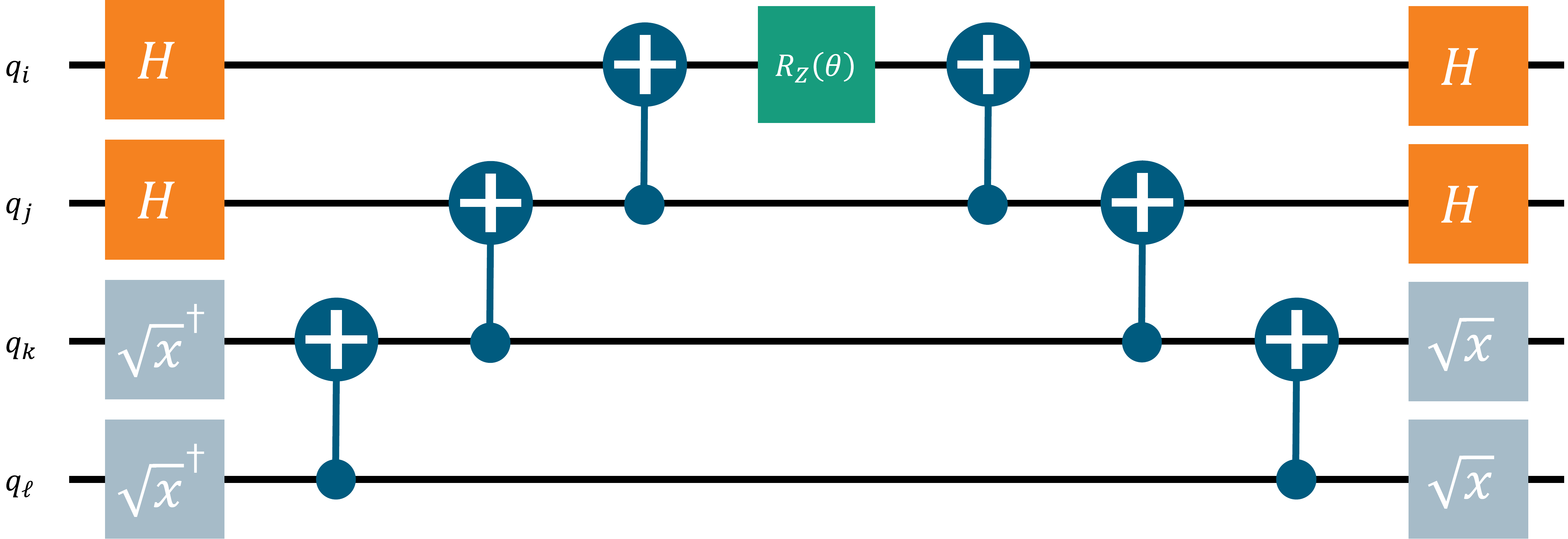}
        \caption{Pauli string $\sigma^x_i \sigma^x_j \sigma^y_k \sigma^y_\ell$ of non-Coulomb two-body term $g_{ijk\ell} a^\dagger_i a^\dagger_j a_k a_\ell + \text{h.c.}$ with the depicted parameter $\theta$ given by $\theta=-g_{ijk\ell} /4$}
        \label{fig:circuits_non_coulomb_two_body}
    \end{subfigure}
  \caption{Circuit representations of the one-body term $h_{ii} a^\dagger_i a_i$ (\subref{fig:circuits_one_body}), the Coulomb two-body term $g_{ijji} a^\dagger_i a^\dagger_j a_j a_i$ (\subref{fig:circuits_couloumb_two_body}), and the Pauli string $g_{ijk\ell}/8 \sigma^x_i \sigma^x_j \sigma^y_k \sigma^y_\ell$ of the non-Coulomb two-body term $g_{ijk\ell} a^\dagger_i a^\dagger_j a_k a_\ell + \text{h.c.}$ (\subref{fig:circuits_non_coulomb_two_body}) using Jordan-Wigner transformation\cite{tranter2018comparisonmapper}. For the non-Coulomb two-body term, the circuit representation of one of the 8 Pauli strings is depicted; the other terms vary solely in the single-qubit rotations prior and after the CNOT ladder structure. For further information about circuit creation from operators, refer to the work by \citeauthor{xu2022decomposition}~\cite{xu2022decomposition}.}
  \label{fig:circuits}
\end{figure}

To identify the bottlenecks in circuit size, it is essential to examine the contributions from one-body and two-body terms.
One-body terms correspond to one-qubit gates (\fref{fig:circuits_one_body}), resulting in a single layer of local one-qubit gates that contribute only a negligible amount to the overall circuit size.
In contrast, the Coulomb two-body terms translate into ZZ gates accompanied by some single-qubit rotations (\fref{fig:circuits_couloumb_two_body}), rendering them relatively efficient for implementation on near-term quantum computers.
However, the topology of the quantum computing hardware can increase circuit size due to the necessity of SWAP gates.
Each non-Coulomb two-body term translates into ladders of CNOT gates, enclosing  parameterized single-qubit Z rotations (\fref{fig:circuits_non_coulomb_two_body}).
Consequently, these terms are highly non-local, requiring all qubits within the system to be interconnected through two-qubit gates.
The number of non-Coulomb two-body terms increases quadratically with the size of the considered molecules. Thus, considering the asymptotic growth of $\mathcal{O}(m^2)$ and the complexity of the circuits, it is evident that the non-Coulomb two-body terms are the bottleneck of \ac{vha} for \ac{nisq} devices.

\subsection{Truncation Scheme} \label{subsec:truncation_scheme}

As noted previously, the one-body terms $H_\alpha$ and the Coulomb two-body terms $H_\beta$ yield small and local quantum circuits (possibly up to SWAP gates due to the hardware topology).
The bottleneck for near-term quantum computers are the non-Coulomb two-body terms contained in $H_\gamma$, introducing large circuits, even for small molecules, that are out of reach for the current quantum computing hardware due to the noise introduced by the vast number of two-qubit gates.
Thus, the proposed truncation scheme is applied solely on $H_\gamma$ keeping all other terms unchanged for circuit construction and consequently the approximation as minimal as possible.

The truncation scheme for \ac{tvha} works as follows.
First, as a preparatory step, the representation of the Hamiltonian $H_\gamma$ is adopted.
By exploiting anticommutation relations, the non-Coulomb two-body terms can be rewritten using the properly antisymmetrized representation with $\tilde g_{ijk\ell} = (g_{ijk\ell}-g_{jik\ell}-g_{ij\ell k}+g_{ji\ell k})$ for $i<j$ and $k<\ell$ and $\tilde g_{ijk\ell} = 0$ for all other cases.
This approach leads to a factor 4 reduction of terms.
This reduction only affects the representation (as a nice side effect reducing memory requirements to store the electronic integrals $ g_{ijk\ell}$) but doesn't cause any difference in the constructed circuit.

Next, the non-Coulomb two-body terms are sorted by their magnitude, i.e. by the size of the $\tilde g_{ijk\ell}$ as they arise in the Hamiltonian $\tilde H_\gamma$.
For simplicity the indices $ijk\ell$ are contracted to the single index $s$ in the sorted sequence of non-Coulomb two-body terms $(\tilde g_{ijk\ell}^\gamma) = (g^\gamma) = (g^\gamma_1, g^\gamma_2, ...)$ with $g^\gamma_r \geq g^\gamma_s$ if $r<s$.
In some cases, there are multiple equal terms $g^\gamma_r = g^\gamma_s$ for $r \neq s$.
In such instances, the order of the terms is determined such that each term is followed by its Hermitian conjugate; apart from this restriction, the order is determined arbitrarily.

As next step, a truncation threshold $p$ is chosen ($0 \leq p \leq 1$), which corresponds to the fraction of non-Coulomb two-body terms included in the circuit construction.
The truncation threshold $p$ is given by 
\begin{equation}
    p = \frac{1}{ \sum_s \left| g^\gamma_s \right| } \sum_{s=1}^{s_{cut}} \left| g^\gamma_s \right|. \label{eq:truncation_threshold}
\end{equation}
$p=0$ represents the limiting case of not taking any non-Coulomb two-body terms into account, i.e. only considering electron correlation based on Coulomb terms.
$p=1$ yields the standard \ac{vha} scheme without any truncation.
As can be seen in \eref{eq:truncation_threshold}, the truncation threshold $p$ can't be selected freely.
Rather, it is dependent on the distribution of the $g^\gamma_s$ terms.
In practical terms, $s_{cut}$ should be chosen in such a way that $p$ approximates the desired truncation threshold as closely as possible.
Values for $s_{cut}$, where a term is separated from its Hermitian conjugate, are forbidden to maintain the Hermitian property of the Hamiltonian.

Rather than using the Hamiltonian $H_\gamma$ which encompasses all non-Coulomb two-body terms, the quantum circuit is eventually constructed using the truncated Hamiltonian $H_\gamma^\text{cut}$ with its terms selected based on the truncation threshold $p$.

Apart from the boundaries of \eref{eq:truncation_threshold} due to the distribution of the $g^\gamma_s$ terms, the truncation threshold can be chosen freely and should be based on the requirements for accuracy on one hand and restrictions due to the noisy hardware to yield small circuits on the other hand.
Thus, the optimal choice of the truncation threshold depends both on the molecule (or more accurately on its Hamiltonian) and the used hardware.
As a rule-of-thumb, a truncation threshold of $p\approx 0.5$ yields a reasonable trade-off between accuracy and circuit size as shown in the following section.

It shall be noted that the truncation is performed on the Fermionic operators.
In principle, it is possible to perform the truncation on the Pauli operators after mapping the Fermionic operators to Pauli spin operators.
While this alternative approach might lead to further reduction in circuit size, it is heavily dependent on the selected transformation scheme (in our case, the Jordan-Wigner mapper) and doesn't allow for a clear physical interpretation of the truncation procedure.

While the truncation scheme is applied to $H_\gamma$ for circuit construction, the complete Hamiltonian is utilized for energy evaluation, specifically for measuring the expectation values of all Pauli words.
A truncation threshold for measurement can be selected, which may significantly reduce the number of required measurements.
Notably, this threshold can differ from the truncation threshold used in the ansatz circuit construction.
The measurement truncation threshold influences only the number of measurements needed, without impacting the circuit properties.
Consequently, it does not alter how well \ac{tvha} circuits perform on \ac{nisq} devices concerning gate errors and coherence time, but it does provide a method for optimizing computation time.
Investigating the application of a measurement truncation threshold for the optimization of this method presents an intriguing area for future research.
Potential inquiries might include the development of criteria for selecting optimal measurement truncation thresholds or evaluating the trade-offs between measurement accuracy and available computational time.

Our \ac{tvha} method can be contrasted with alternative approaches for reducing the operator set for circuit construction, such as \ac{adaptvqe}\cite{grimsley2019adaptive} and its multiple variants like Qubit-ADAPT-VQE~\cite{tang2021qubit,dalton2024quantifying}.
The \ac{adaptvqe} method grows the circuit iteratively, which demands significantly more computation time.
On the other hand, \ac{tvha} employs a straightforward truncation scheme to select operators, which greatly speeds up the ansatz creation.
This efficiency stems from \ac{tvha}'s selection of operators from \ac{vha}, which has access to information about the operator magnitude in the chemical system.
Conversely, \ac{adaptvqe} typically creates its operator pool from excitations resembling the \ac{ucc} ansatz or other operator pools (e.g. pools of Pauli words with specific symmetry constraints). Both ways have in common that they lack information about the expected magnitude of the operators.
This necessitates an extensive iterative loop to evaluate which operators to include.
As a result, \ac{tvha} can rapidly achieve relatively small circuits, while \ac{adaptvqe} can produce close-to-optimal circuits but at the cost of significantly increased computation time.

%% file: 4_section_results_and_discussion.tex
\section{Results of Numerical Studies} \label{sec:results_of_numerical_calculations}

To assess our proposed method, we evaluate the calculated energy, the CNOT count, and the number of parameters for a variety of molecular systems, comparing the results to the state-of-the-art ansatze \ac{ucc} with single and double excitations (\ac{uccsd}), and additional triple excitations (\ac{uccsdt}), as well as \ac{hea}.
The energy deviation from \ac{fci} provides a measure of the method's accuracy.
The CNOT count serves as an indicator of applicability on \ac{nisq} hardware, as the number of CNOT gates that can be executed on quantum hardware is limited by the error of the 2-qubit gates.
This can to some extent be mitigated with error mitigation schemes like \ac{zne}\cite{kim2023scalable}.
The circuit depth, which is restricted by the decoherence time of the hardware device, could also be considered.
However, since circuit depth and CNOT count are closely related, we chose to use the CNOT count as the evaluation metric.
The parameters count completes the evaluation metrics, indicating how well the classical optimization routine is expected to perform, especially in the context of scaling to larger system sizes.

We have selected the molecular systems \ac{lih}, \ac{h2}, \ac{h4}, and \ac{ch2} for our benchmark to ensure a diverse range of systems.
\ac{lih}, the most complex system considered with the highest requirements regarding computation metrics like number of needed qubits, serves as the primary system.
Its complexity level is high enough to effectively assess the performance of \ac{tvha}, yet it remains manageable both on classical hardware for the reference \ac{fci} energy calculation, and for executing the proposed quantum algorithm on a simulator of a quantum computer on classical hardware.
Additionally, the hydrogen molecule, a standard benchmark in this field, is included to demonstrate the applicability of \ac{tvha} in such typically studied systems.
The hydrogen chain presents a well-known system characterized by highly correlated electrons, offering a more complex scenario for evaluating the performance of \ac{tvha}.
Finally, methylene exemplifies a system with significant electron correlation, demonstrating the applicability of \ac{tvha} in conjunction with active space calculations, thereby highlighting its versatility in tackling challenging molecular scenarios.

The calculations were performed with the Qiskit\cite{qiskit2024} framework utilizing their implementation of \ac{vqe} as well as the SBPLX\cite{NLopt,SUBPLEX} optimizer.
The Jordan-Wigner mapper is used due to its easy interpretability; for application-based calculations beyond scientific studies other mappers might be preferable and can easily be exchanged with the Jordan-Wigner mapper.
Refer to~\cite{tranter2018comparisonmapper} for a detailed comparison of possible mappers.

All calculations are performed using Qiskit's statevector simulator, resembling a 'perfect' quantum computer without shot noise nor any other noise source.
This allows to evaluate the proposed algorithm and benchmark it to state-of-the-art algorithms without additional uncontrollable sources of noise due to the still heavily fluctuating performance of \ac{nisq} hardware, enabling a fair comparison of the ansatze.
It shall be emphasized that the aim of this paper is to compare these algorithms and not to benchmark the day-dependent performance of quantum hardware.
For our claim of \ac{nisq} compatibility serve the above mentioned metrics of CNOT count and number of parameters within the ansatz.

As reference serve \ac{fci} calculations (in case of \ac{ch2} \ac{casci} calculations using a (2,~2) active space) and an allowed tolerance of 1.5\,mHa in accordance with the convention of chemical accuracy.
Unless stated otherwise, the electronic energies are given; the overall energies of the molecules can be obtained by adding the nuclear repulsion energy which is a precomputed constant given by each molecule's geometry; for \ac{ch2}, the per-computed energy of the inactive space must also be added to obtain the overall energy.
Please refer to \sref{app:sec:molecule_setup} for details on the molecule setup (including the basis sets used) and to \sref{app:sec:numerical_calculations} as well as our published code\cite{possel2025tvha} for further information on the quantum simulations.

\subsection{Lithium Hydride}

\begin{figure}[!htb]
  \centering
  \includegraphics[width=0.8\textwidth]{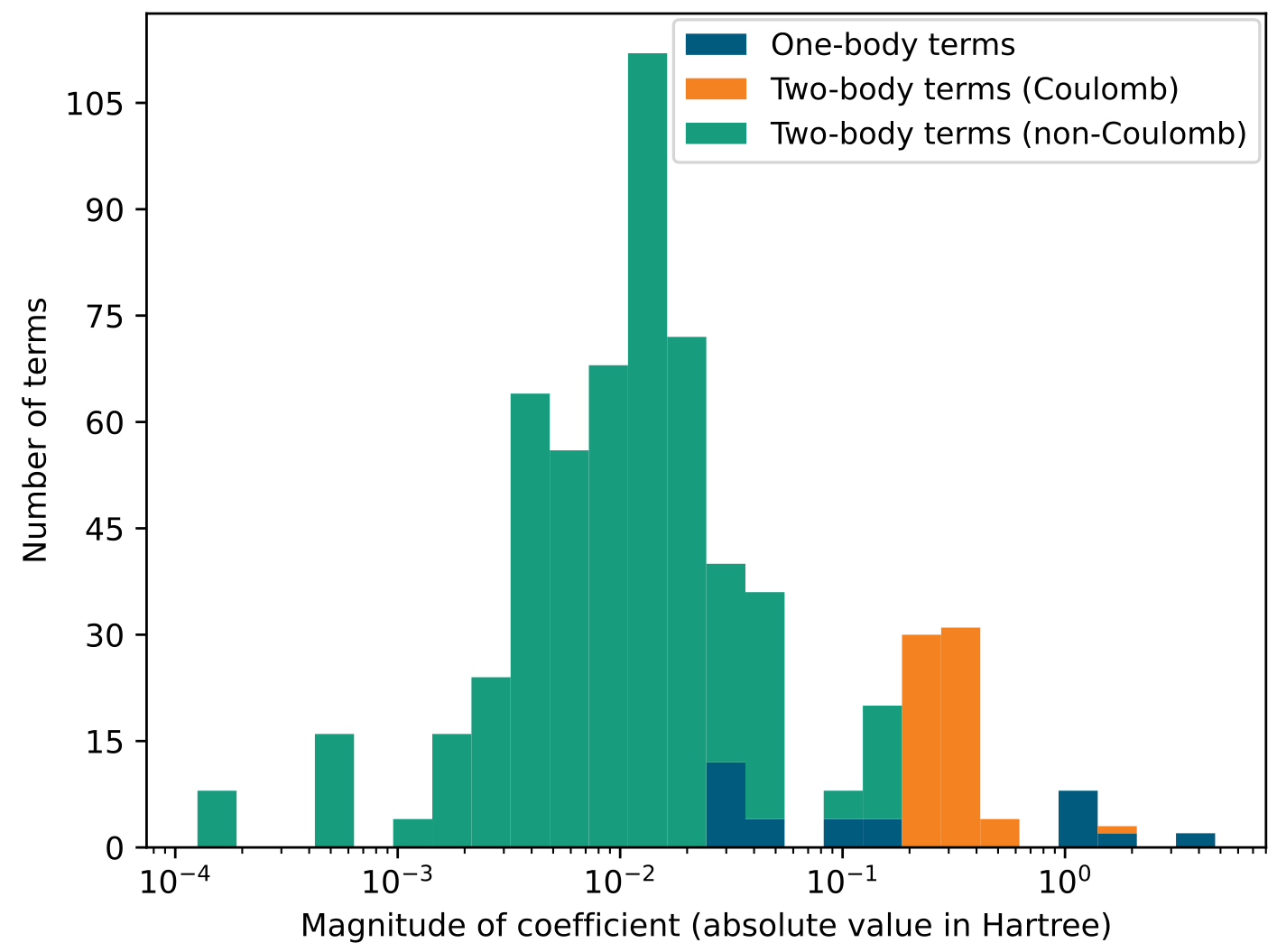}
  \caption{Distribution of the magnitude of one-body terms $h_{ii}$ and (Coulomb and non-Coulomb) two-body terms $g_{ijk\ell}$ of the lithium hydride molecule. Note: the x-axis is displayed on a logarithmic scale, reflecting the large differences in order of magnitude among the terms.}
  \label{fig:hist_lih}
\end{figure}

In our study, we describe \ac{lih} as a system with 12 spin orbitals, translating to 12 qubits on a quantum computer.
As such, \ac{lih} provides a system that is classically fully tractable with the chosen \ac{fci} method and basis set. And yet it is sufficiently complex to evaluate the effectiveness of the \ac{tvha}.
More details about the molecule system can be found in the supplementary information.

The distribution of one-body and two-body terms in the Hamiltonian of the LiH molecule (see \fref{fig:hist_lih}) shows that the terms are distributed across various orders of magnitude. The non-Coulomb two-body terms are (up to a few exception) by far the smallest terms, which provides additional justification to the chosen truncation scheme acting on the non-Coulomb two-body terms.

\begin{figure}[!htb]
  \centering
  \includegraphics[width=0.8\textwidth]{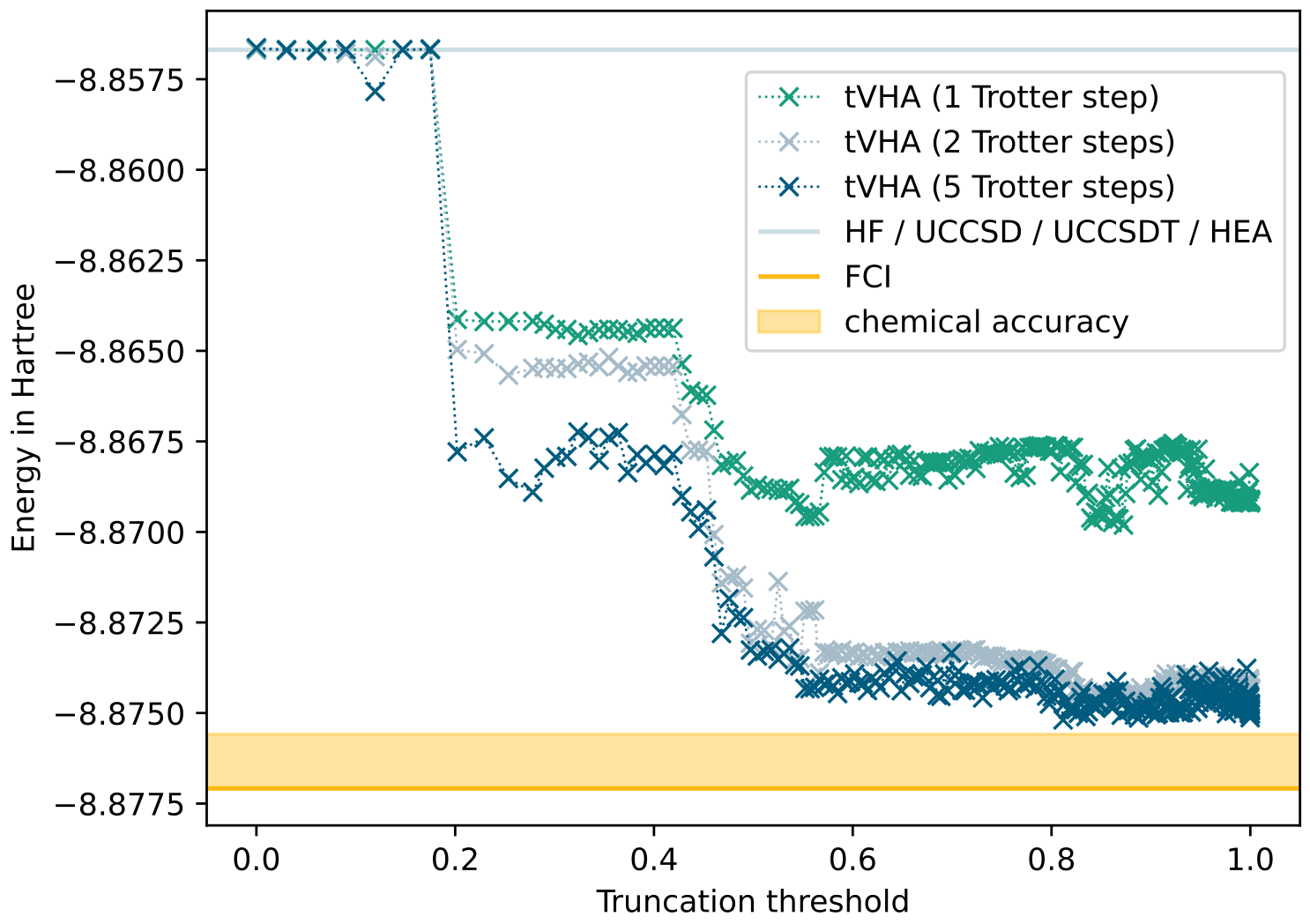}
  \caption{Calculated energy of the lithium hydride molecule dependent on the truncation threshold $p$.
  Depicted is the energy calculated with \ac{tvha} with 1, 2, and 5 Trotter steps, respectively, as well as the energies obtained with \ac{uccsd} and \ac{hea}.
  For comparison, \ac{hf} energy and \ac{fci} with the chemical accuracy are given.
  }
  \label{fig:energy_threshold_lih}
\end{figure}

Due to the large number of non-Coulomb two-body terms, the energy plot (\fref{fig:energy_threshold_lih}) shows multiple features.
A truncation threshold of $p \approx 0.2$ is already sufficient to include a certain amount of electron correlation but doesn't allow to reach chemical accuracy yet.
The energy accuracy can be further improved by adding more terms until $p \approx 0.5$; afterwards the energy accuracy doesn't change significantly, in other words, the added two-body terms don't introduce further information about the system.
Thus, these terms can be considered redundant.
Even the \ac{vha} border case of $p=1$ including all terms doesn't suffice reaching the ground state within chemical accuracy.
This can be explained with the limitation of a single Trotter step.
Here, the variational approach is not capable to fully compensate the Trotterization error.
So, additional Trotter steps are required.

\begin{figure}[!htb]
    \centering
    \includegraphics[width=0.8\textwidth]{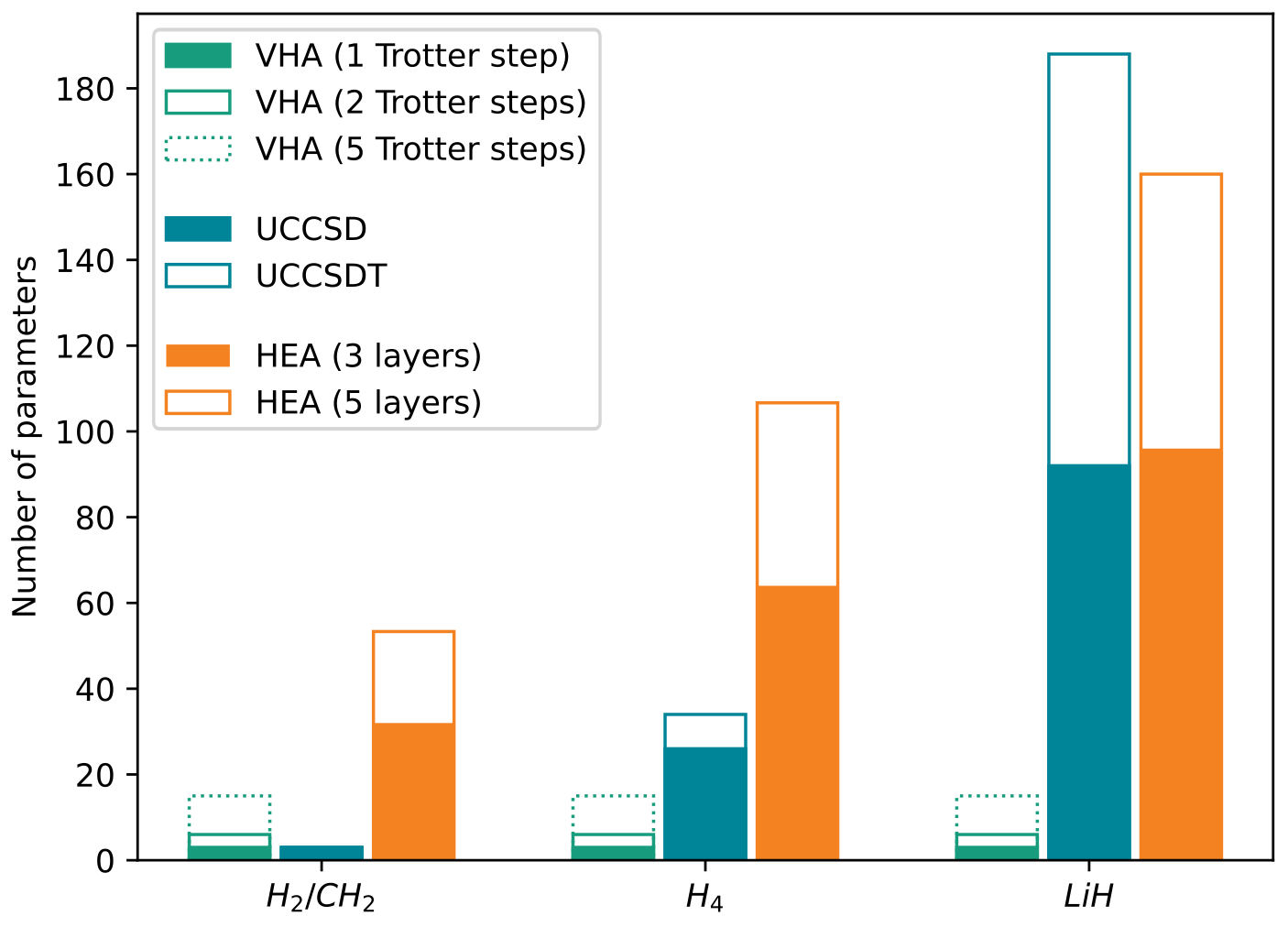}
    \caption{Parameter count of \ac{vha}, \ac{ucc} and \ac{hea} for the molecules \ac{h2}, \ac{ch2}, \ac{h4}, and \ac{lih}.
    To depict the scaling behaviour of the ansatze, the respective systematic improvements are shown, too:
    For \ac{vha}, the number of Trotter steps is increased, for \ac{ucc} the list of excitations (singles and doubles, additionally triples), for \ac{hea} the number of repeated layers (3 layers are used throughout the paper, 5 layers are shown here for demonstration).} 
    \label{fig:parameter_count}
\end{figure}

In principle it is possible to add more parameters to the ansatz within a single Trotter layer.
With this approach, one would basically get closer to the idea behind the \ac{ucc} ansatz, which parametrizes every single excitation available, to make the ansatz more expressive and to try and further compensate for the Trotterization error.
However, this approach comes with its own limitations, mainly a highly increased number of parameters (see also \fref{fig:parameter_count}) that becomes increasingly complicated to optimize as can be clearly seen in \fref{fig:energy_threshold_lih}, where the \ac{sbplx} optimizer was not able to escape the local minimum of the \ac{hf} solution in case of the highly-parameterized ansatz \ac{uccsd} (nor for \ac{hea}).
So, the \ac{tvha} focuses on keeping the number of parameters low at the expensive of adding more Trotter steps and thus increasing the circuit size linearly with the number of Trotter steps.

\Fref{fig:energy_threshold_lih} shows that the critical point of $p \approx 0.2$ is stable for the different numbers of Trotter steps, consolidating our claim that the non-Coulomb two-body terms below this threshold aren't sufficient to represent electronic correlations.
Also the step at $p \approx 0.5$ of improved energy accuracy is stable for all numbers of Trotter steps.
Adding a second Trotter step is sufficient to get as close as 3\,mHa to the \ac{fci} solution---still twice the allowed chemical accuracy of 1.5\,mHa.

There are two possible explanations for the ansatz's failure to achieve chemical accuracy with $p=1$ and as much as 5 Trotter steps.
The first explanation suggests that even 5 Trotter steps may be insufficient. 
However, the relative stability of energy predictions with two or more Trotter steps counters this notion.
The second explanation considers the details about the hybrid variational algorithm, which has a classical optimization step after each quantum mechanical measurements.
The depicted results are achieved with the classical \ac{sbplx} optimizer with an allowed maximum of 1000 function evaluations.
It is likely that the optimizer fails to locate the global optimum (or any other optimum within chemical accuracy) within the specified number of function evaluations.
The number of parameters grows linearly with the number of Trotter steps but the maximum number of function evaluations remains constant in our framework.
Further investigations using different optimizers as well as adjusting the allowed number of function evaluations could elucidate this issue.

\begin{figure}[!htb]
  \centering
  \includegraphics[width=0.8\textwidth]{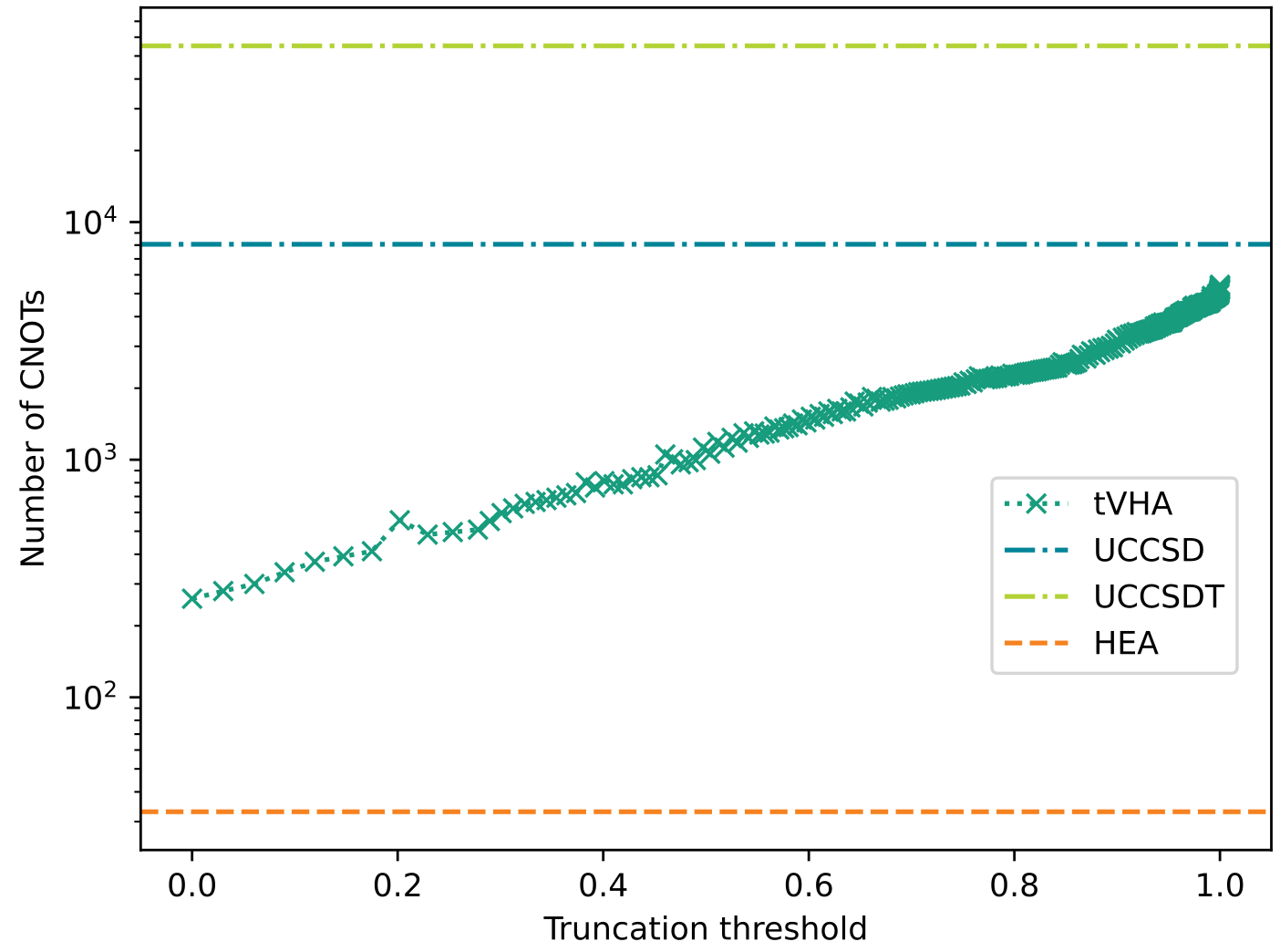}
  \caption{Number of non-local gates (CNOT gates) of the quantum circuit created with \ac{tvha} of the lithium hydride molecule dependent on the truncation threshold $p$ for a single Trotter step.
  For comparison, the CNOT count for the circuits created with \ac{uccsd}, \ac{uccsdt}, and \ac{hea} are shown.}
  \label{fig:cnot_count_lih}
\end{figure}

Analysis of the CNOT count (see \fref{fig:cnot_count_lih}) reveals significant improvements for calculations on \ac{nisq} devices: the \ac{vha} approach (equals \ac{tvha} with $p=1$) necessitates over 5000 CNOT gates on an ideal all-to-all connected device. However, by employing a truncation threshold of $p=0.5$, the CNOT count is reduced by a factor of 5. Considering that 2 Trotter steps are required, this results in approximately 2000 CNOT gates for \ac{tvha}, yielding an overall reduction factor of 2.5 compared to standard \ac{vha} while still maintaining high accuracy in energy prediction.
For comparison, the CNOT count for \ac{hea} and \ac{ucc} is presented.
As expected, \ac{hea} requires the smallest number of CNOTs, thus demonstrating its hardware efficiency.
The CNOT count of \ac{uccsd} is comparable to that of \ac{vha} without any truncation.
When examining the systematic improvements of each method, the advantages of \ac{tvha} become evident:
it necessitates more Trotter steps, resulting in a linear increase in the CNOT count with the number of Trotter steps.
In contrast, \ac{ucc} extends to triple excitations, significantly increasing the CNOT count (not to mention the further increase with quadruple excitations).
There have been efforts in the scientific community to enhance \ac{ucc} without incorporating triple excitations by adding additional repetitions of the circuit.
This can be viewed as a hybrid ansatz that attempts to integrate concepts from the adiabatic theorem into the excitation-based \ac{cc} approach.
\ac{vha}, on the other hand, offers a mathematically more consistent way to improve the ansatz with a linear increase in circuit size.
Lastly, \ac{hea} can be improved by adding more layers; however, the other methods converge in the theoretical limits of an infinite number of Trotter steps (i.e., the adiabatic limit as 
$t\rightarrow \infty $ for \ac{vha}) and infinite excitations (or, more concisely, all available excitations, i.e., \ac{fci} for \ac{ucc}).
In contrast, the improvement of \ac{hea} does not, to the authors' knowledge, guarantee that one can even sample the appropriate region of the Hilbert space where the solution to the problem resides.

Side note on connectivity and concrete hardware devices: All-to-all connectivity is only available in certain hardware architectures, such as ion trap-based quantum computers. In other architectures, the number of CNOT gates may be significantly higher, depending on the connectivity topology and the transpilation routine employed.
Additionally, the native two-qubit gate may vary across hardware implementations, such as Møller-Plesset gates on devices based on Rydberg atoms.

\clearpage

\subsection{Hydrogen Molecule}

\begin{figure}[!htb]
  \centering
  \includegraphics[width=0.8\textwidth]{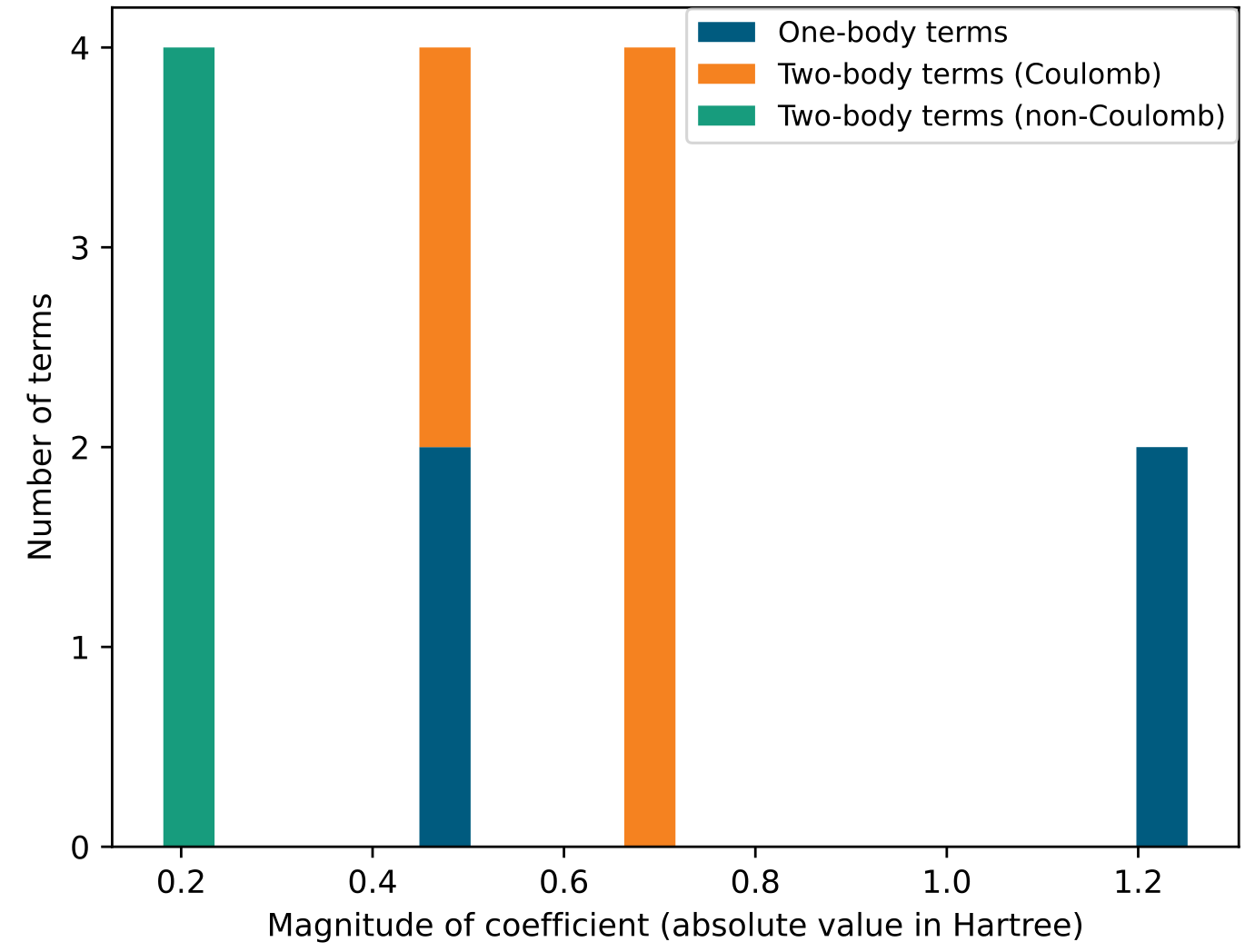}
  \caption{Distribution of the magnitude of one-body terms $h_{ii}$ and (Coulomb and non-Coulomb) two-body terms $g_{ijk\ell}$ of the hydrogen molecule.}
  \label{fig:hist_h_2}
\end{figure}

The hydrogen molecule, a fundamental system widely used in quantum chemistry applications on quantum computing platforms, serves as another example. 
Its simplicity doesn't underscore the concept of \ac{tvha} as effectively as the lithium hydride example, but it still provides a basis for demonstrating that the \ac{tvha} is applicable even in such basic systems.
Its electronic Hamiltonian describes a system of two electrons in two spatial orbitals, i.e. four spin orbitals.
Utilizing the Jordan-Wigner mapper, this system can be represented by Pauli operators (Pauli words of size 4), which can be translated to a quantum circuit with 4 qubits.
Symmetry reduction, utilizing the spatial symmetry of the \ac{h2} molecule, would allow to form a trivial 2-qubit system; this simplification was left out on purpose as the resulting system would be too trivial to gain reasonable insights about \ac{tvha}.
The distribution of one-body and two-body terms with respect to their magnitude in the Hamiltonian can be seen in \fref{fig:hist_h_2}.
As expected, the non-Coulomb two-body terms are significantly smaller than the Coulomb two-body and the one-body terms; due to the overall small number of terms, the asymptotically quadratic number of two-body terms compared to the linear number of one-body terms is not visible in this example.
Thus, the expected bottleneck of the vast amount of two-body terms, that is clearly visible in \ac{lih} (see \fref{fig:hist_lih}) and other larger systems, is not apparent in this small system of \ac{h2}.

\begin{figure}[!htb]
  \centering
  \includegraphics[width=0.8\textwidth]{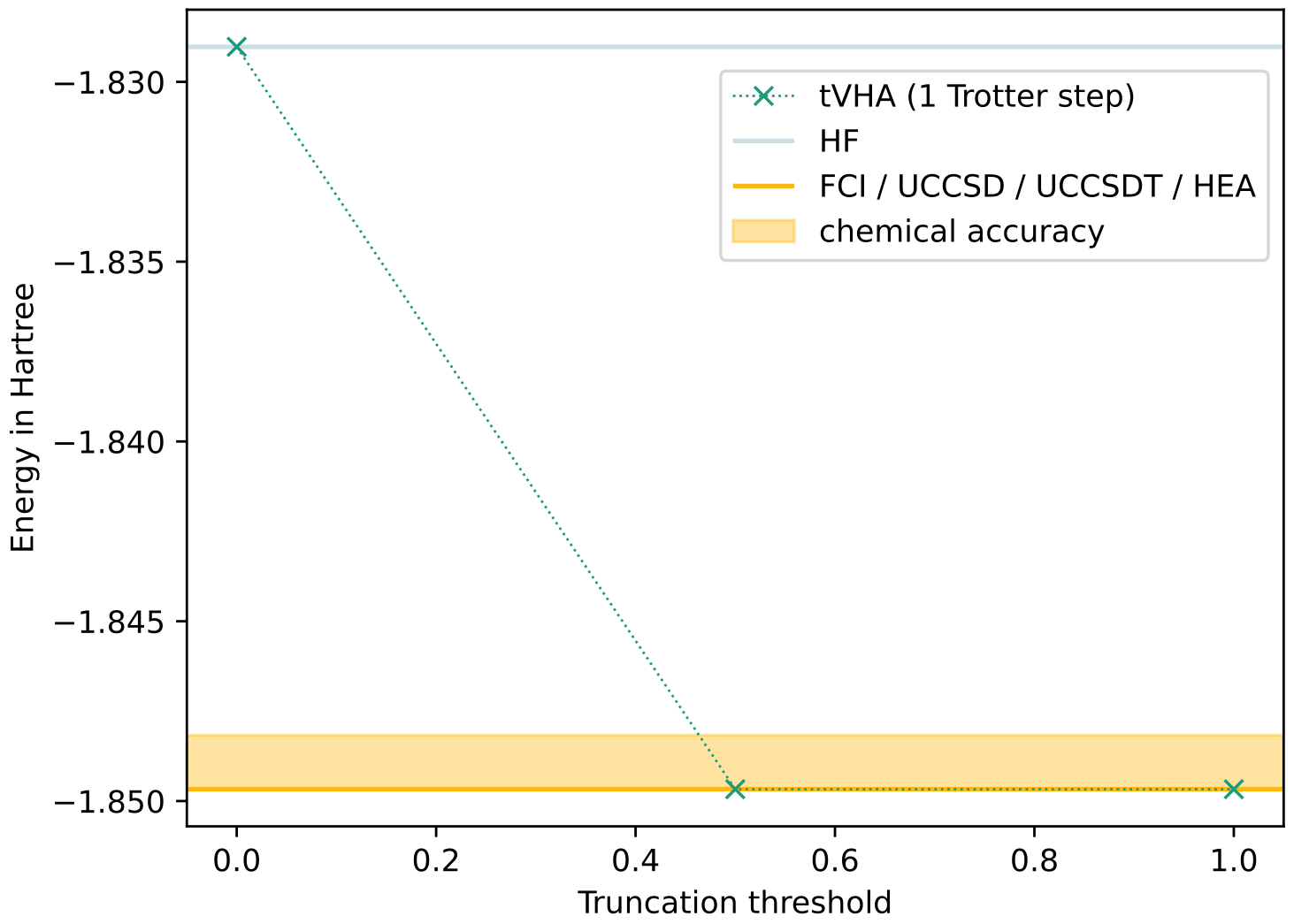}
  \caption{Calculated energy of the hydrogen molecule dependent on the truncation threshold $p$.
  Depicted is the energy calculated with \ac{tvha} with 1 Trotter steps, as well as the energies obtained with \ac{uccsd} and \ac{hea}.
  The calculated energies for 2 and 5 Trotter steps are identical to those for 1 Trotter step and are therefore omitted from the plot.
  In this specific representation of the \ac{h2} molecule, there are no triple excitations; hence, \ac{uccsdt} is equivalent to \ac{uccsd}.
  For comparison, \ac{hf} energy and \ac{fci} with the chemical accuracy are given.
  Due to the simplicity of the \ac{h2} molecule and its limited number of non-Coulomb two-body terms (see \fref{fig:hist_h_2}), only three distinct truncation thresholds exist.
  }
  \label{fig:energy_threshold_h_2}
\end{figure}

Since the hydrogen molecule is such a simple system, already a single Trotterization step is sufficient to reach chemical accuracy for standard \ac{vha} as can be seen in \fref{fig:energy_threshold_h_2} with a truncation threshold $p=1$.
Also a threshold of $p=0.5$ is sufficient, neglecting half of the non-Coulomb two-body terms.
Neglecting all non-Coulomb two-body terms with $p=0$ prevents the ansatz to cover electron correlations and thus does not allow any improvement over the \ac{hf} solution.
It can be concluded that some non-Coulomb two-body terms are necessary to cover electron correlation but there is a certain redundancy within the non-Coulomb two-body terms, allowing to remove those and adjust the parameters of the other terms in the variational approach to still be able to capture the physics of the system and reach the ground state energy.
As shown in \fref{fig:energy_threshold_h_2}, both \ac{uccsd} and \ac{hea} are capable of incorporating electron correlation as well.

\begin{figure}[!htb]
  \centering
  \includegraphics[width=0.8\textwidth]{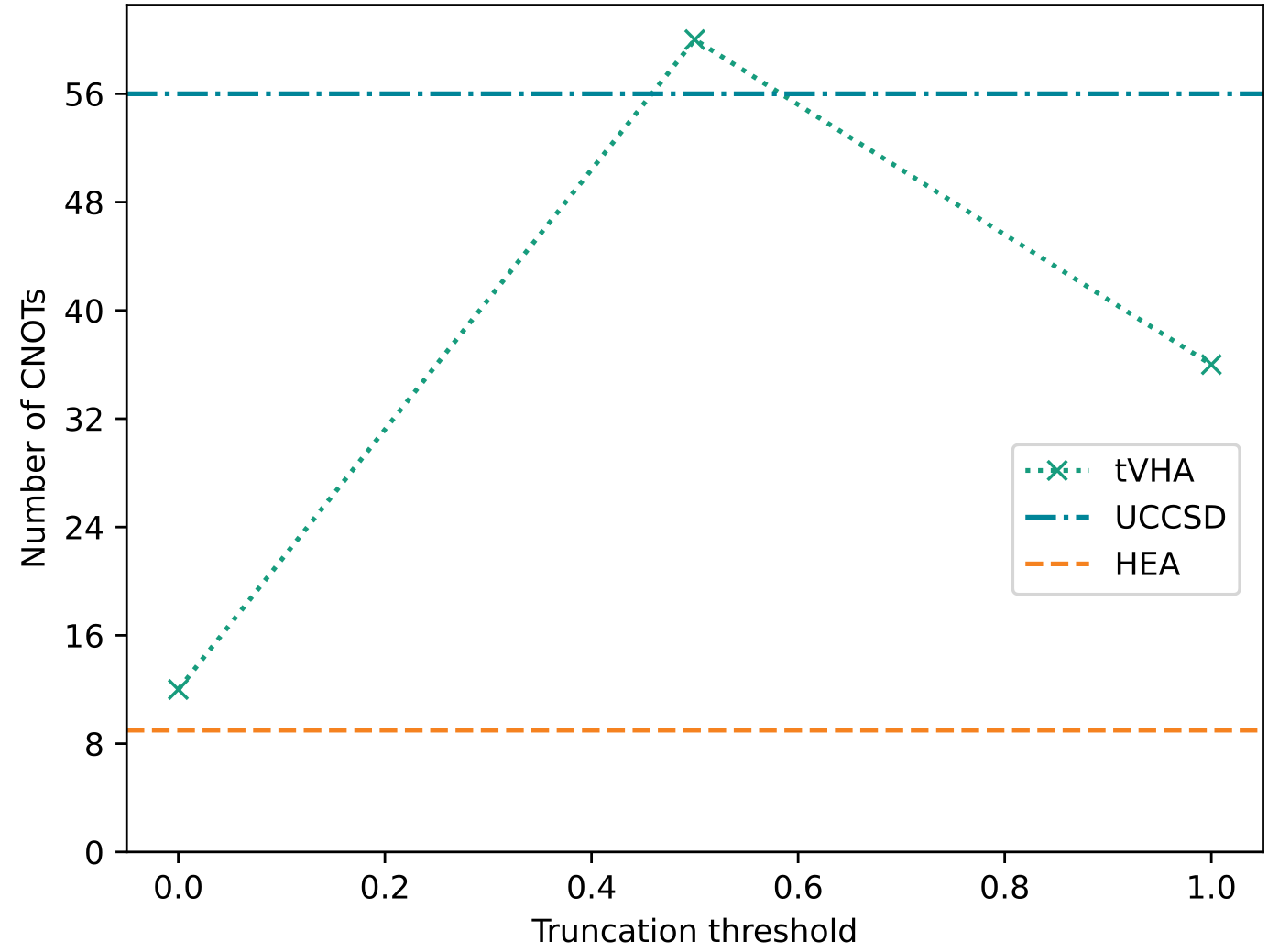}
  \caption{Number of non-local gates (CNOT gates) of the quantum circuit created with \ac{tvha} of the hydrogen molecule dependent on the truncation threshold $p$.
  For comparison, the CNOT count for the circuits created with \ac{uccsd} and \ac{hea} are shown.
  In this specific representation of the \ac{h2} molecule, there are no triple excitations; hence, \ac{uccsdt} is equivalent to \ac{uccsd}.
  Due to the simplicity of the \ac{h2} molecule and its limited number of non-Coulomb two-body terms (see \fref{fig:hist_h_2}), only three distinct truncation thresholds exist.
  }
  \label{fig:cnot_count_h_2}
\end{figure}

Looking at the constructed quantum circuits for these calculations, one can see an artifact in the hydrogen molecule (see \fref{fig:cnot_count_h_2}.
While the number of CNOT gates (in our work equivalent to the number of two-qubit gates) and the overall circuit depth increase from $p=0$ to $p=1$ as expected, $p=0.5$ is a special case.
The hydrogen molecule exhibits a high degree of spatial symmetry. When translating the Fermionic operators into Pauli operators using the Jordan-Wigner mapper, some Pauli operators cancel each other out, resulting in smaller circuits.
However, for $p=0.5$, the spatial symmetry is compromised due to the removal of certain Fermionic operators.
Consequently, there are no cancellations among the Pauli operators, leading to a larger quantum circuit.
This artifact can, in principle, occur in any molecule exhibiting symmetries.
In this academic case, where symmetry reduction was intentionally omitted, the artifact became particularly evident.
In more practical scenarios, one would typically apply symmetry reduction first and then implement \ac{tvha} on the resulting system, thereby avoiding this artifact.

\clearpage

\subsection{Hydrogen Chain}

\begin{figure}[!htb]
  \centering
  \includegraphics[width=0.8\textwidth]{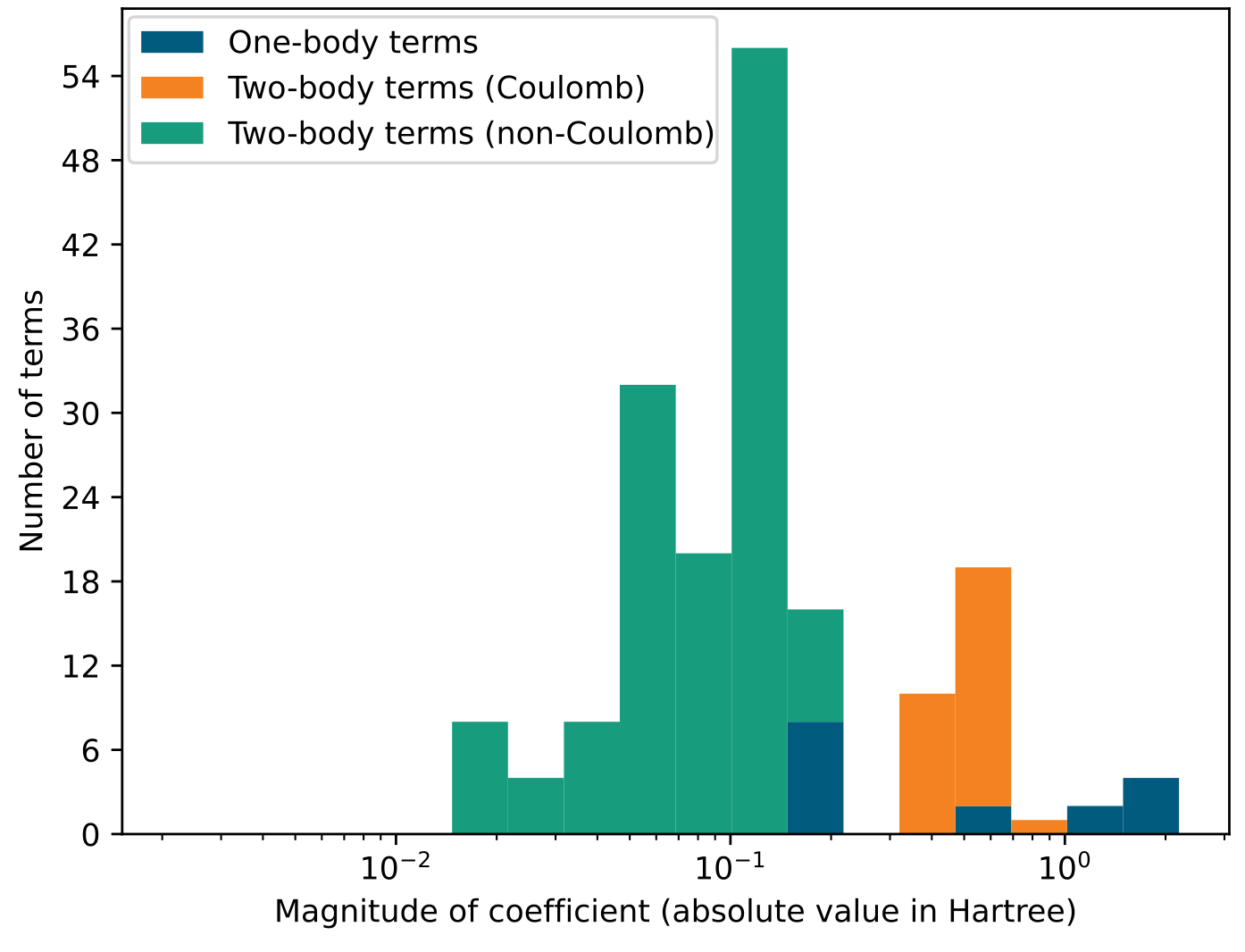}
  \caption{Distribution of the magnitude of one-body terms $h_{ii}$ and (Coulomb and non-Coulomb) two-body terms $g_{ijk\ell}$ of the hydrogen chain.
  Note: the x-axis is displayed on a logarithmic scale, reflecting the large differences in order of magnitude among the terms.}
  \label{fig:hist_h_4}
\end{figure}

The linear hydrogen chain of four atoms, with equal distances between atoms matching the equilibrium distance in the hydrogen molecule, serves as an example of systems with highly correlated electrons.
Despite its smaller size of 8 spin orbitals compared to the 12 spin orbitals of \ac{lih}, its distribution of terms (see \fref{fig:hist_h_4}) exhibits a similar pattern to that of \ac{lih}.

\begin{figure}[!htb]
  \centering
  \includegraphics[width=0.8\textwidth]{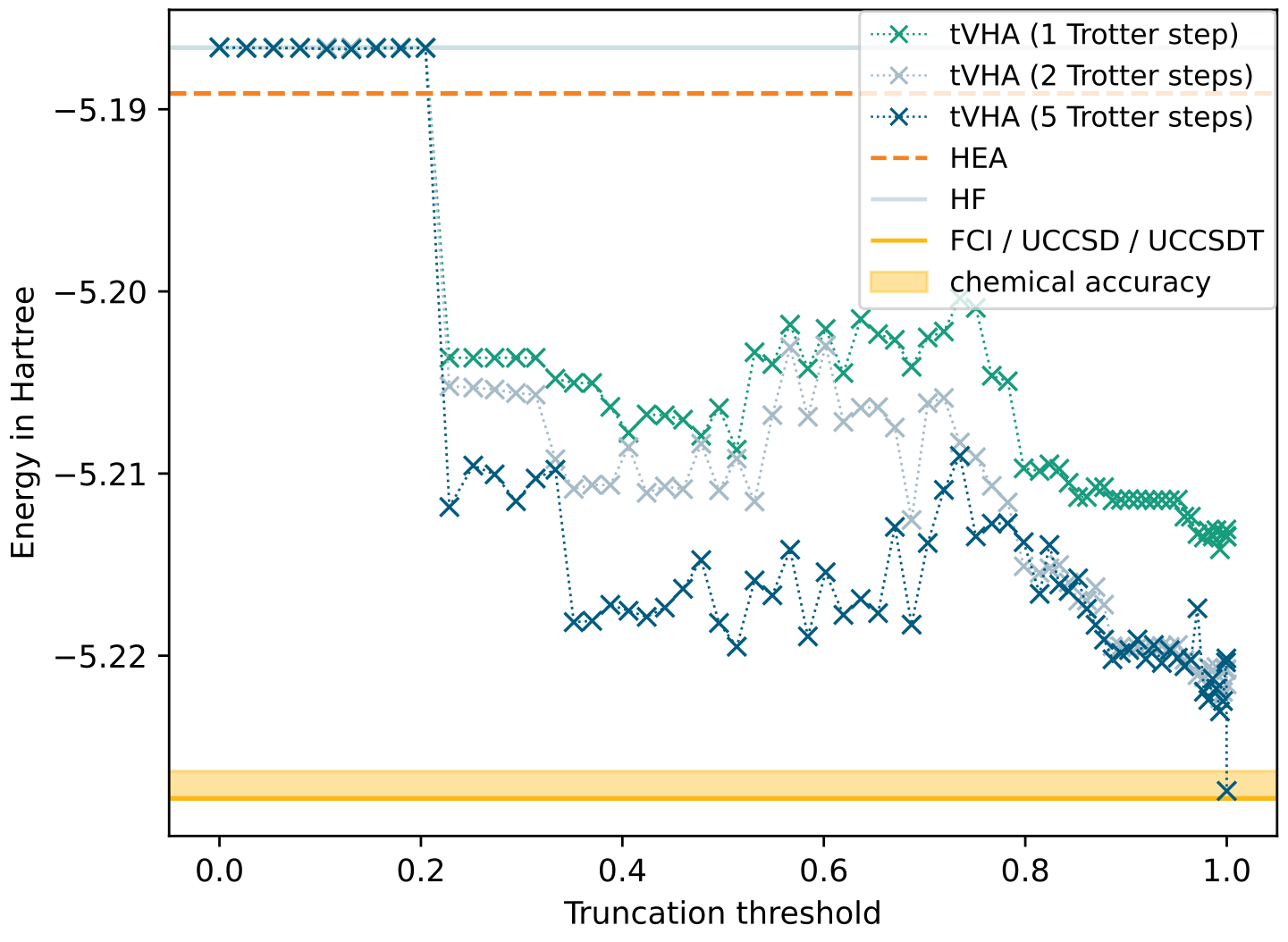}
  \caption{Calculated energy of the hydrogen chain dependent on the truncation threshold $p$.
  Depicted is the energy calculated with \ac{tvha} with 1, 2, and 5 Trotter steps, respectively, as well as the energies obtained with \ac{uccsd} and \ac{hea}.
  For comparison, \ac{hf} energy and \ac{fci} with the chemical accuracy are given.
  }
  \label{fig:energy_threshold_h_4}
\end{figure}

The dependence of the calculated energy of \ac{h4} on the truncation threshold (\fref{fig:energy_threshold_h_4}) also exhibits features similar to those of \ac{lih}.
Electron correlations start to be incorporated around a truncation threshold of 0.2, with an intermediate optimum observed around 0.5, which is close to the best result at $p=1$.
However, this value still falls considerably short of achieving chemical accuracy.
\Fref{fig:energy_threshold_h_4} furthermore demonstrates that these features remain stable with an increased number of Trotter steps, while overall, additional Trotter steps significantly improve the results, bringing them closer to chemical accuracy.
Further increasing the number of Trotter steps does not enhance accuracy.
This phenomenon highlights the artifacts introduced by the optimizer and its insufficient number of function evaluations, as previously discussed in the context of lithium hydride.
In comparison, the calculation using \ac{hea} does not achieve chemical accuracy; however, the optimizer is able to escape the local minimum provided by the \ac{hf} solution.
In contrast, both \ac{uccsd} and \ac{uccsdt} are sufficient to attain chemical accuracy.

\begin{figure}[!htb]
  \centering
  \includegraphics[width=0.8\textwidth]{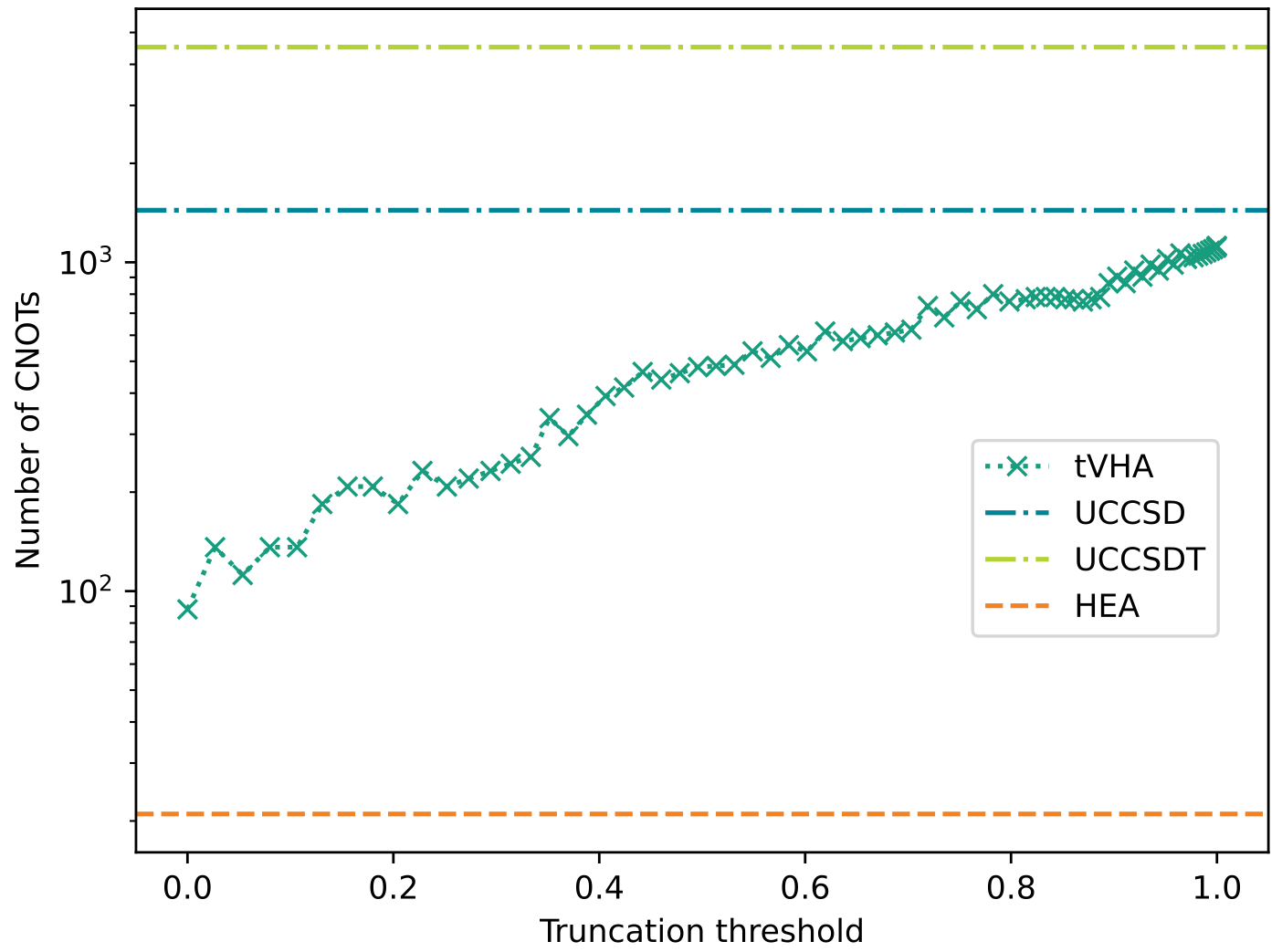}
  \caption{Number of non-local gates (CNOT gates) of the quantum circuit created with \ac{tvha} of the hydrogen chain dependent on the truncation threshold $p$ for a single Trotter step.
  For comparison, the CNOT count for the circuits created with \ac{uccsd}, \ac{uccsdt}, and \ac{hea} are shown.}
  \label{fig:cnot_count_h_4}
\end{figure}

In the hydrogen chain, the CNOT count exceeds 1100 for the full \ac{vha} approach, thus being slightly smaller than the \ac{uccsd} ansatz, while it decreases to less than 500 for \ac{tvha} with a truncation threshold around $0.5$, representing an improvement by roughly a factor of 2.2 (see \fref{fig:cnot_count_h_4}.
This substantial reduction in CNOT count enhances the usability of \ac{tvha}, making it more feasible for practical implementations in quantum computing while still maintaining adequate accuracy for the system under consideration as opposed to \ac{hea} which (though efficient regarding the CNOT count) failed for the hydrogen chain to make significant improvements over the \ac{hf} solution.
It is noteworthy that \ac{tvha} can efficiently address systems with highly correlated electrons, as demonstrated by this example of a hydrogen chain.
Furthermore, if one is willing to compromise on accuracy, it is possible to further decrease the CNOT count, potentially improving efficiency in resource-constrained environments.

\clearpage

\subsection{Methylene}

\begin{figure}[!htb]
  \centering
  \includegraphics[width=0.7\textwidth]{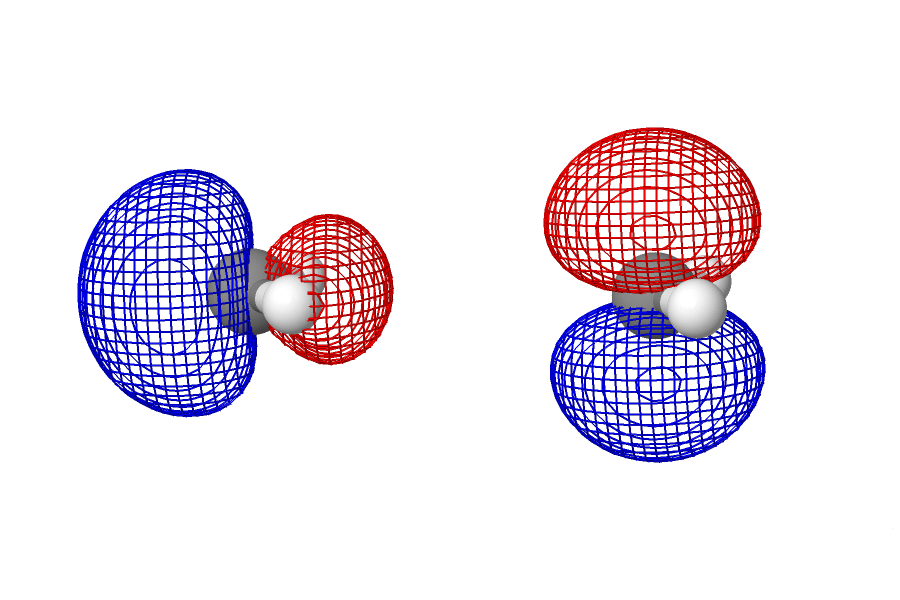}
  \caption{Active orbitals of \ac{ch2}.}
  \label{fig:active_orbitals_ch_2}
\end{figure}

\begin{figure}[!htb]
  \centering
  \includegraphics[width=0.8\textwidth]{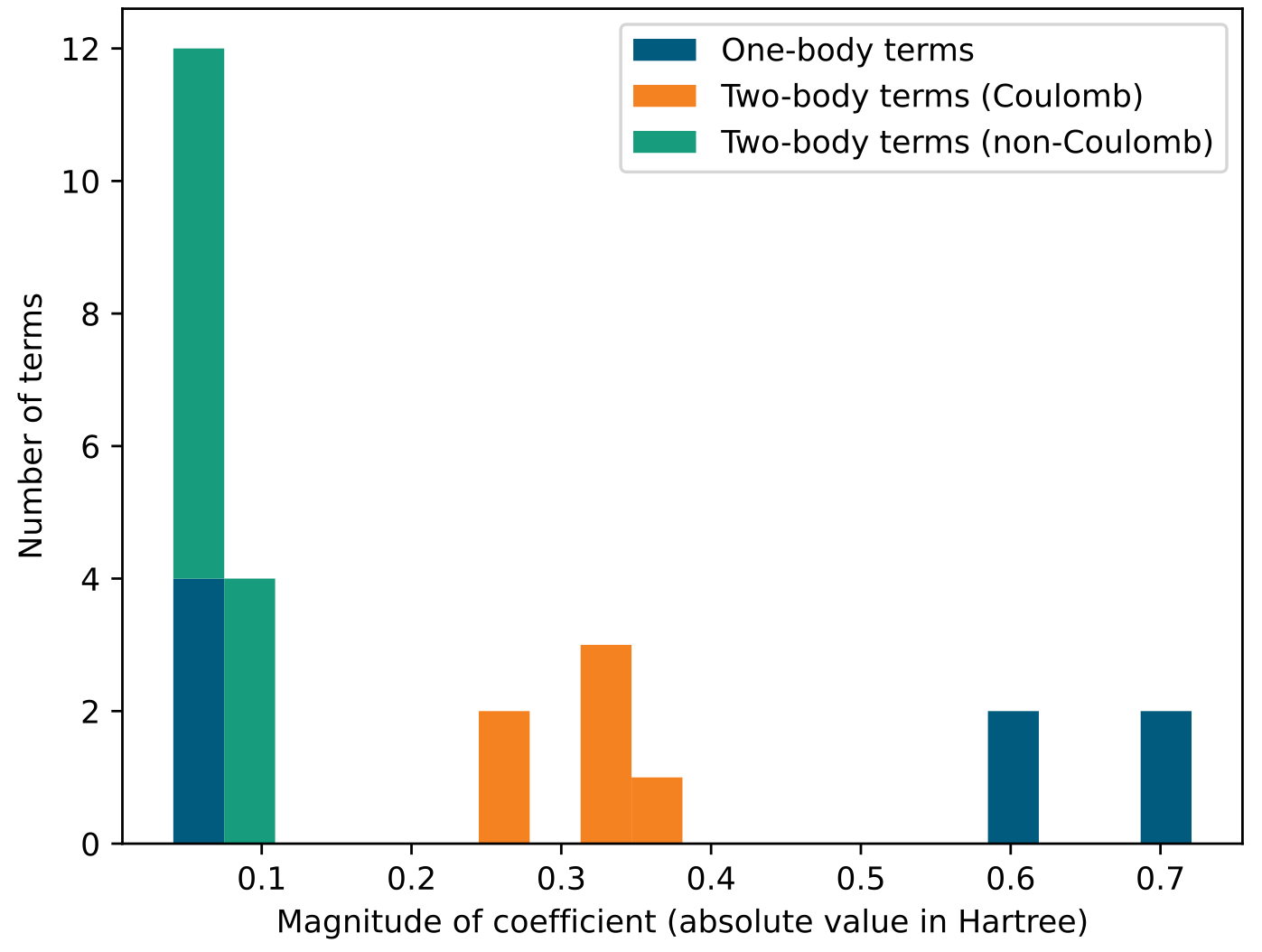}
  \caption{Distribution of the magnitude of one-body terms $h_{ii}$ and (Coulomb and non-Coulomb) two-body terms $g_{ijk\ell}$ of the methylene molecule.}
  \label{fig:hist_ch_2}
\end{figure}

While the previous molecules were evaluated using their full orbital space, the methylene example utilizes active space reduction through \ac{asf}, showing that \ac{tvha} is fully compatible with active space methods.
The selected active space of two spatial orbitals (see \fref{fig:active_orbitals_ch_2}) results in a representation using four qubits on the quantum computer.
It requires the same number of qubits as the hydrogen molecule; however, it exhibits a less trivial distribution of one-body and two-body terms (see \fref{fig:hist_ch_2}), along with reduced symmetry compared to hydrogen.

\begin{figure}[!htb]
  \centering
  \includegraphics[width=0.8\textwidth]{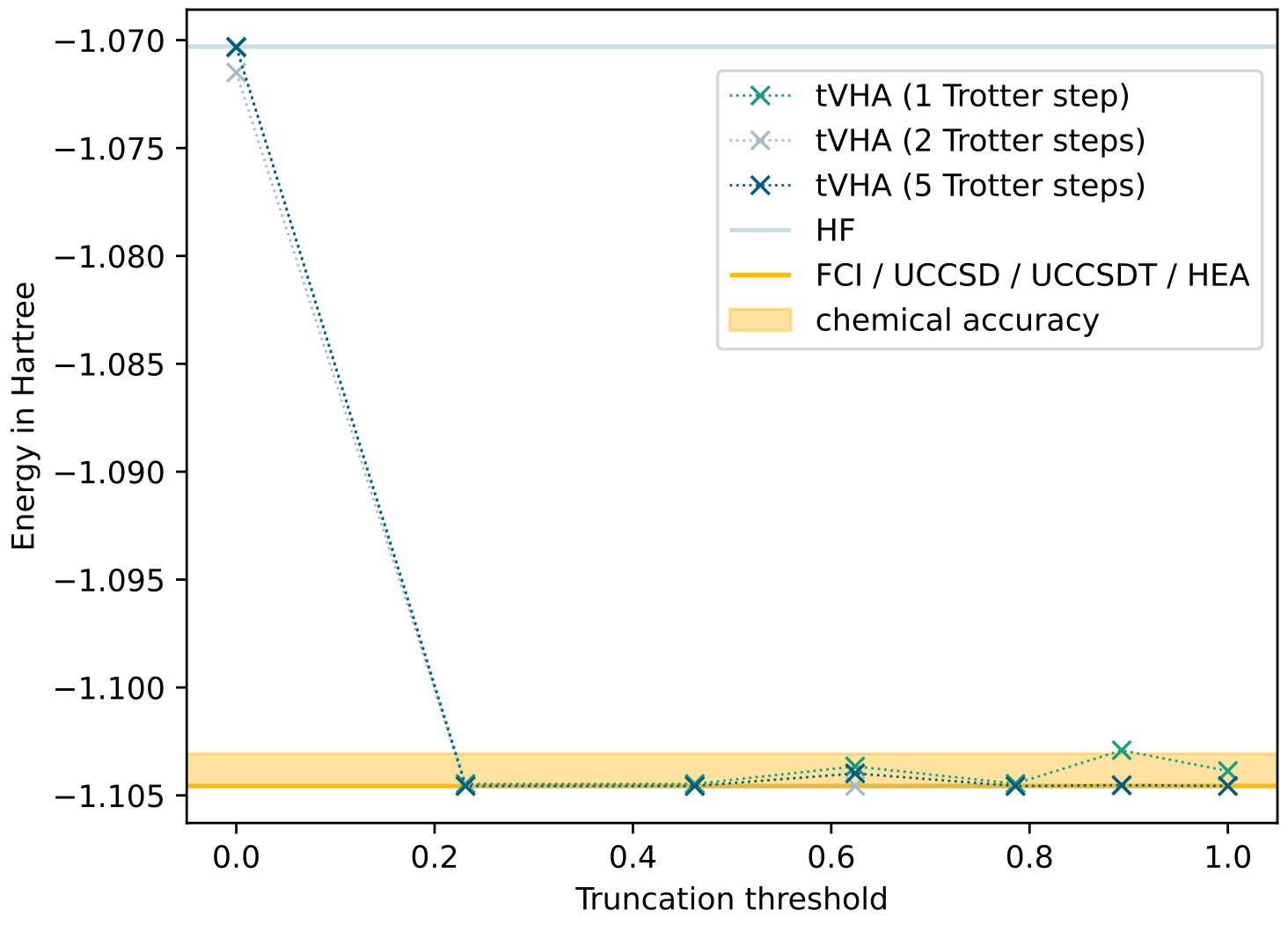}
  \caption{Calculated energy of the active space of the methylene molecule dependent on the truncation threshold $p$.
  Depicted is the energy calculated with \ac{tvha} with 1, 2, and 5 Trotter steps, respectively, as well as the energies obtained with \ac{uccsd} and \ac{hea}.
  For comparison, \ac{hf} energy and \ac{fci} with the chemical accuracy are given.
  }
  \label{fig:energy_threshold_ch_2}
\end{figure}

\begin{figure}[!htb]
  \centering
  \includegraphics[width=0.8\textwidth]{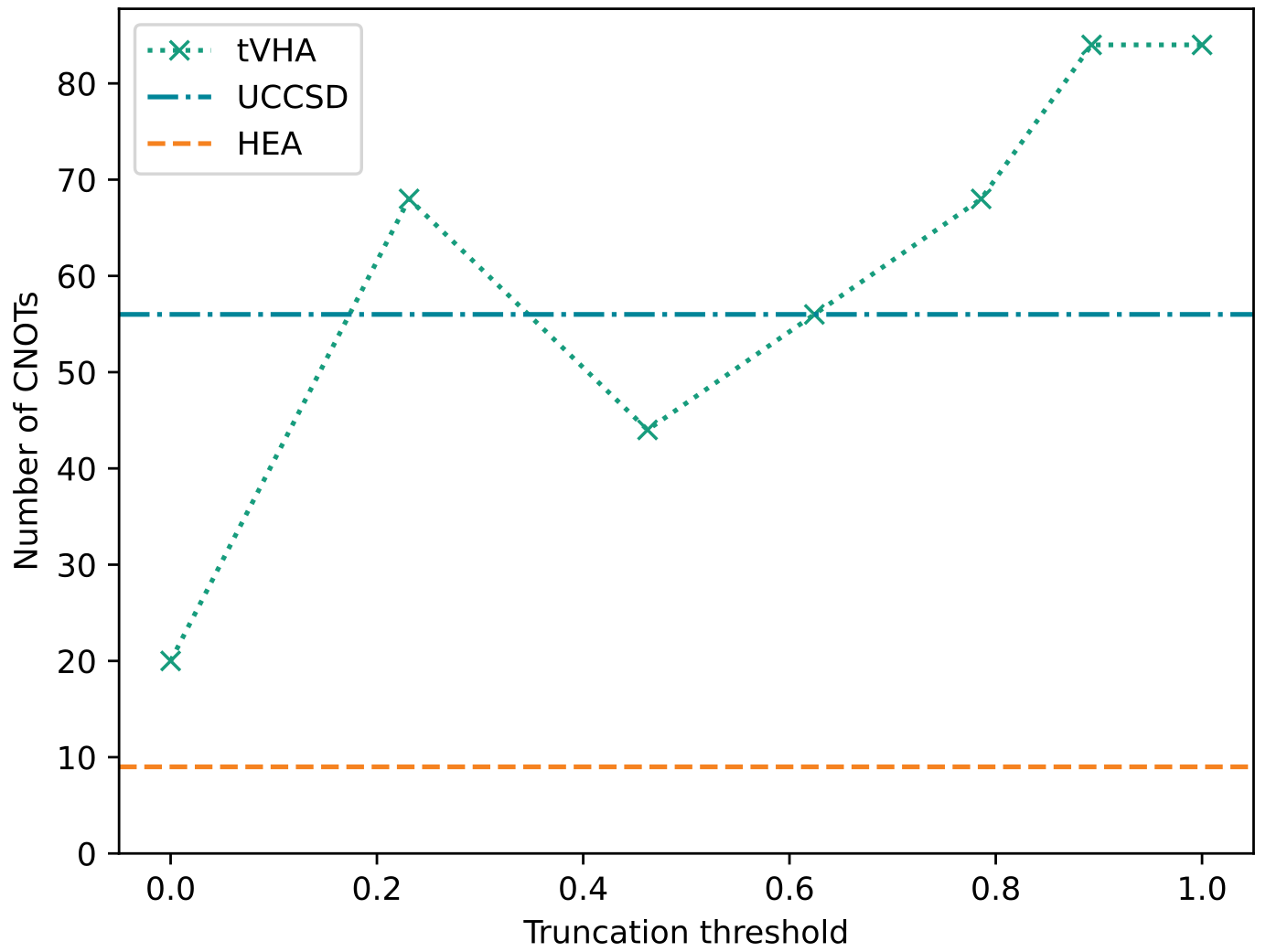}
  \caption{Number of non-local gates (CNOT gates) of the quantum circuit created with \ac{tvha} of the methylene molecule dependent on the truncation threshold $p$ for a single Trotter step.
  For comparison, the CNOT count for the circuits created with \ac{uccsd} and \ac{hea} are shown.
  \ac{uccsdt} is omitted since there are no triple excitations within the chosen active space of \ac{ch2}.
  }
  \label{fig:cnot_count_ch_2}
\end{figure}

\Fref{fig:energy_threshold_ch_2} demonstrates that a single Trotter step and a truncation threshold above $0.2$ is already sufficient to reach chemical accuracy within the active space, making \ac{tvha} a highly efficient choice for the calculation of \ac{ch2}.
Although \ac{ch2} does not possess the same level of symmetry as \ac{h2}, it still displays certain symmetric characteristics, as illustrated by the artifact in \fref{fig:cnot_count_ch_2}: Utilizing only a single non-Coulomb two-body term results in a larger circuit than incorporating a second term.
This occurs because the symmetry allows some of the Pauli terms to cancel each other out following the Jordan-Wigner transformation. Nevertheless, even with only one non-Coulomb two-body term, there is a reduction in circuit size without reducing the accuracy; with an optimal truncation threshold around 0.5, this reduction can reach a factor of 2.

Thus, the \ac{ch2} system demonstrates that \ac{tvha} can be applied for active space calculations of highly correlated systems.

\clearpage

%% file: 5_section_summary_and_outlook.tex
\section{Summary and Outlook}\label{sec:summary_and_outlook}

In this research, we introduced and evaluated \ac{tvha}, a novel quantum circuit design for conducting quantum calculations on \ac{nisq} devices.
We demonstrated that our proposed \ac{tvha} significantly reduces the parameter count and can potentially decrease circuit size substantially, making quantum calculations more feasible on \ac{nisq} devices.
The key findings from our study are:

Efficient Circuit Design: tVHA leverages principles from the adiabatic theorem in solid state physics to design efficient quantum circuits.
The proposed ansatz introduces a novel truncation scheme that optimizes the operators used in circuit construction to further reduce circuit depth with minimal classical computation overhead,  outperforming existing methods.

Significant Reduction in Parameter and Circuit Counts: Our proposed \ac{tvha} significantly reduces the parameter count compared to state-of-the-art ansätze such as \ac{uccsd} and \ac{hea}.
This results in easier convergence within the variational quantum eigensolver framework, facilitating quantum calculations on \ac{nisq} devices.

Balancing Accuracy and Efficiency: \ac{tvha} aims to balance accuracy and efficiency.
The introduced truncation scheme allows for the inclusion of some electron correlation, while excluding redundant information.
This approach strikes a balance between the necessity of incorporating electron correlation and the desire for computational efficiency.

Compatibility with Active Space Calculations: We demonstrated that \ac{tvha} can be easily combined with active space methods.
This was exemplified with the methylene molecule, indicating the versatility of \ac{tvha} in addressing challenging molecular scenarios involving highly correlated electrons and active space calculations.

Our future work will focus on developing a scheme to automatically predict the best truncation threshold and number of Trotterization steps based on hardware restrictions, which will allow for a more efficient balancing of hardware requirements and algorithmic accuracy.
We also aim to investigate the possibility to perform the truncation procedure on the Pauli operators to further reduce the circuit size.
It may also be beneficial to explore an adaptive scheme rooted in the adiabatic theorem, which would create the operator pool and ansatz design in a manner akin to \ac{tvha}.
This approach would incorporate an iterative loop similar to existing \ac{adaptvqe} implementations, but with a crucial distinction: it would allow for a sophisticated preselection of operators that is not feasible with \ac{ucc}-based operator pools.
Additionally, benchmarks of \ac{tvha} against alternative methods like \ac{adaptvqe}, \ac{uccsd}, and \ac{hea} on real quantum computing hardware could provide further insights.
However, this would currently require relatively small systems due to the limitations of existing \ac{nisq} hardware.
We anticipate that with the continuous advancement in quantum hardware, it will be possible to test our approach on larger, more complex systems both from quantum chemistry and material science in the near future.

In conclusion, our work presents \ac{tvha} as a promising approach for conducting quantum calculations on \ac{nisq} devices.
While this paper concentrated on the practical applications of \ac{tvha} in quantum chemistry, its underlying principles suggest a wider applicability, extending to the broader field of material science computations on quantum computing platforms.
We believe our work paves the way for further research into the development of efficient quantum algorithms for quantum chemistry and material science.

%% file: acronyms.tex
\begin{acronym}
    \acro{asf}[ASF]{Active Space Finder}
    \acro{adaptvqe}[AdaptVQE]{Adaptive Variational Quantum Eigensolver}
    \acro{ch2}[CH\textsubscript{2}]{methylene}
    \acro{casci}[CASCI]{Complete Active Space Configuration Interaction}
    \acro{casscf}[CASSCF]{Complete Active Space Self-Consistent Field}
    \acro{cc}[CC]{Coupled Cluster}
    \acro{ci}[CI]{Configuration Interaction}
    \acro{dev2-svp}[dev2-SVP]{Development 2\textsuperscript{nd} generation - Split Valence Polarization}
    \acro{dft}[DFT]{Density-Functional Theory}
    \acro{dmrg}[DMRG]{Density-Matrix Renormalization Group}
    \acro{hf}[HF]{Hartree-Fock}
    \acro{h2}[H\textsubscript{2}]{hydrogen molecule}
    \acro{h4}[H\textsubscript{4}]{hydrogen chain}
    \acro{fci}[FCI]{Full Configuration Interaction}
    \acro{hea}[HEA]{Hardware-Efficient Ansatz}
    \acro{hva}[HVA]{Hamiltonian Variational Ansatz}
    \acro{lih}[LiH]{lithium hydride}
    \acro{mp2}[MP2]{second-order M{\o}ller-Plesset perturbation theory}
    \acro{mp}[MP]{M{\o}ller-Plesset}
    \acro{nisq}[NISQ]{Noisy Intermediate-Scale Quantum}
    \acro{pyscf}[PySCF]{Python-based Simulations of Chemistry Framework}
    \acro{qaoa}[QAOA]{Quantum Approximate Optimization Algorithm}
    \acro{qpe}[QPE]{Quantum Phase Estimation}
    \acro{sbplx}[SBPLX]{Subplex}
    \acro{scf}[SCF]{self-consistent field}
    \acro{slsqp}[SLSQP]{Sequential Least Squares Quadratic Programming}
    \acro{spsa}[SPSA]{Simultaneous perturbation stochastic approximation}
    \acro{sto3g}[STO-3G]{Slater-type orbitals with 3 Gaussian functions}
    \acro{tvha}[tVHA]{truncated Variational Hamiltonian Ansatz}
    \acro{ucc}[UCC]{Unitary Coupled Cluster}
    \acro{uccsd}[UCCSD]{Unitary Coupled Cluster with Single and Double excitations}
    \acro{uccsdt}[UCCSDT]{Unitary Coupled Cluster with Single, Double, and Triple excitations}
    \acro{vha}[VHA]{Variational Hamiltonian Ansatz}
    \acro{vqa}[VQA]{Variational Quantum Algorithm}
    \acro{vqe}[VQE]{Variational Quantum Eigensolver}
    \acro{zne}[ZNE]{Zero Noise Extrapolation}
\end{acronym}

%% file: acknowledgements.tex
\section*{Acknowledgements}

This research was supported by funding from the Ministerium für Wirtschaft, Arbeit und Tourismus Baden-Württemberg (Ministry of Economic Affairs, Labour and Tourism of Baden-Württemberg) through the projects QC-4-BW, QC-4-BW II, and KQCBW24. 

The work of DB was partially funded by the Horizon Europe programme HORIZON-CL4-2022-QUANTUM-01-SGA via the
project 101113946 OpenSuperQPlus100.

%% file: conflict_of_interests.tex
\section*{Conflict of Interests}

The authors declare that there are no competing interests.

%% file: supplementary_information.tex
\appendix\section*{Appendix}

\section{Molecule Setup} \label{app:sec:molecule_setup}

For \ac{lih} we have estimated the equilibrium distance with \ac{pyscf}\cite{pyscf2018} and the basis set \ac{sto3g} to be $1.596\,\text{\AA}$.
For \ac{h2} the equilibrium distance of $0.74279\,\text{\AA}$ is chosen based on the same type of calculations.
 \ac{ch2}, structure was optimized for the triplet ground state using B3LYP\cite{becke1993density}\cite{lee1988development}\cite{stephens1994ab} Def2-TZVP \cite{weigend2005balanced} functional and D3BJ dispersion corrections. The optimization calculation was performed using ORCA 5 \cite{neese2022software} Frequencies were also calculated and no imaginary frequencies were observed.  \tref{tab:ch2_geometry}. 
\begin{table}[htb]
    \centering
    \begin{tabular}{lrrr}
        C &  -0.00000000558058 & 0.0 &  0.27482875331272 \\
        H & 0.99659455942822 & 0.0 & -0.13738437780928 \\
        H & -0.99659455384764 & 0.0 &  -0.13738437550343 
    \end{tabular}
    \caption{Equilibrium geometry of \ac{ch2}.}
    \label{tab:ch2_geometry}
\end{table}

For \ac{h4} no geometry optimization is performed; instead the equilibrium distance of \ac{h2} is used as distance between each pair of the linear chain of hydrogen molecules.

All molecules are represented in the \ac{sto3g} basis set except for \ac{ch2}, for which the \ac{dev2-svp} basis set is used.

\subsection{Active space selection}

For \ac{ch2} the minimal the two singly occupied of the triplet state (sp2 and p) orbitals are chosen as these orbitals are nearly degenerate \cite{ghafarian2019spin}\cite{ghafarian2018accurate}.

\section{Numerical Calculations} \label{app:sec:numerical_calculations}

The numerical calculatons were performed with Qiskit\cite{qiskit2024} version 1.4.2 and its sub-packages, including qiskit-nature version 0.7.2.
The \textit{NumPyMinimumEigensolver} of qiskit-algorithms was used to perform numerical diagonalization to calculate the \ac{fci} energies.
Our package tVHA\cite{possel2025tvha} was used for \ac{tvha} calculations and creation for the plots.

For the classical optimization part of the \ac{vqe} loop, the optimizer \ac{sbplx} was used due to its favorable properties.
Although it typically requires more iterations for convergence than gradient-based optimizers like \ac{slsqp}, it shows good resilience against small amounts of errors as arise, for example, in simulations including shot noise. While typical gradient-based optimizers fail under any (even small) noise levels, gradient-agnostic algorithms like \ac{spsa} or the Nelder-Mead algorithm handle these environments better, though they come with other shortages, mostly significantly slower convergence behaviour or even failure of convergence.
As a generalization of the Nelder-Mead algorithm, \ac{sbplx} shows acceptable convergence speed and high fidelity to reach the global (or at least a sufficiently good) minimum, both in noise-free and noisy environments, and is thus suitable both for statevector simulations and for noisy hardware.
In this study, the number of function evaluations of the \ac{sbplx} algorithm was restricted to 1\,000; while systems with few parameters easily convergerce within this time, this number of function evaluations is not sufficient to ensure convergence for larger parameter spaces with more than 15 parameters (i.e., more than 5 Trotter steps).
Since classical optimization is a wide and complex field, this study is restricted to a single well-working  optimization algorithm without further hyperparameter optimization, on purpose leaving out fine-tuning of the optimization routine and the optimizer's hyperparameters which would go well beyond the scope of this paper.

The error sources that can arise during a quantum computing calculation are the following 
\begin{itemize}
    \item device noise: e.g. depolarization noise due to unwanted physical interactions with the environment of the hardware; below a certain noise threshold it can in principle be suppressed with arbitrary precision with error correction codes
    \item shot noise: due to the stochastic character of quantum measurements, the underlying probability distribution of a wavefunction can only be estimated; within the limit of an infinite number of shots/measurements this error can be statistically suppressed
    \item optimization errors: in general, convergence to the global ground state of a parameter space cannot be guaranteed; however, a large amount of iterations/function evaluations and careful choice of the optimization algorithm and its hyperparameters typically allows to reach minima close to the global minimum as long as no Barren plateaus arise
    \item algorithmic errors: these errors stem from assumptions and approximations within the algorithm for calculation of the basic properties, here the \ac{tvha} with its error sources coming from Suzuki-Trotter approximation, discretization of the time evolution, and truncation of small non-Coulomb two-body terms
    
\end{itemize}
The statevector simulation allows to track the system's wavefunction throughout the whole circuit evaluation and thus resembles a perfect quantum computer (i.e. absence of device noise) within the limit of an infinite number of shots (i.e. absence of shot noise).
Thus, the simulations performed in this study allow for the deepest insight into the proposed algorithm itself, minimizing all other sources of errors.

The reference calculations for \ac{uccsd} and \ac{uccsdt} were performed using Qiskit routines, specifically the \ac{ucc} class with the \ac{hf} initial state.
The resulting circuits resemble those of \ac{tvha}, as both utilize single and double excitations, or equivalently, one-body and two-body terms.
However, the \ac{ucc} circuits contain significantly more free parameters than \ac{tvha}.

\begin{figure}[htb]
    \centering
    \includegraphics[width=0.95\textwidth]{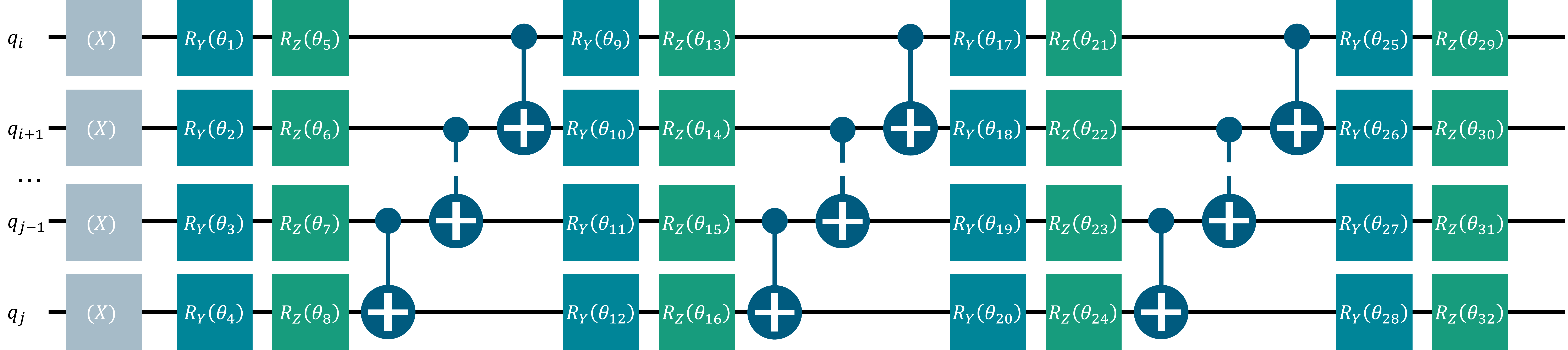}
    \caption{Circuit of \ac{hea} with the \ac{hf} initial state.
    The figure depicts our custom \ac{hea} circuit with 3 layers based on Qiskit's EfficientSU2 ansatz.
    The brackets around the NOT gates (represented as 'X' gates) indicate that their application depends on the number of layers and qubits required to reproduce the \ac{hf} state when all parameters are set to zero.
    For \ac{h2}, all 4 qubits are preceded by a NOT gate.
    }
    \label{fig:si_hea_with_hf_initialstate}
\end{figure}

For the reference calculations of \ac{hea}, a custom class based on Qiskit's EfficientSU2 ansatz was implemented. Since Qiskit's implementation of the EfficientSU2 ansatz does not support the \ac{hf} initial state, we added this functionality by inserting NOT gates. This modification ensures that, after circuit evaluation with all parameters set to zero, the \ac{hf} state is produced. The overhead of this initialization method is at most one NOT gate per qubit, making it negligible.
For the rotation layer, parameterized single-qubit $R_Y$ rotations are followed by parameterized $R_Z$ rotations.
The subsequent entanglement layer is constructed using the 'reverse linear' entanglement scheme.
In total, three repetitions or layers are employed in our calculations, where a single layer consists of an $R_Y$ layer, an $R_Z$ layer, and an entanglement layer.
The circuit layout is also depicted in \fref{fig:si_hea_with_hf_initialstate}.

